# Magnetic Monopoles and Duality Symmetry Breaking in Maxwell's Electrodynamics

## Jay R. Yablon[*]

910 Northumberland Drive

Schenectady, New York, 12309-2814

**Abstract:**

It is shown how to break the symmetry of a Lagrangian with duality symmetry between electric and magnetic monopoles, so that at low energy, electric monopole interactions continue to be observed but magnetic monopole interactions become very highly suppressed to the point of effectively vanishing. The "zero-charge" problem of source-free electrodynamics is solved by requiring invariance under continuous, local, duality transformations, while local duality symmetry combined with local $U(1)_{EM}$ gauge symmetry leads naturally and surprisingly to an $SU(2)_D$ duality gauge group.

---

[*] jyablon@nycap.rr.com

# 1. Introduction

The question of whether magnetic monopoles exist in nature, and if so, in what form, has been one of nature's deepest mysteries almost since the time of Maxwell, and remains so today.[*] The Maxwell-duality formalism, first proposed by Reinich [1] and later elaborated by Wheeler, [2] (we shall refer to this as "Reinich-Wheeler Duality"), provides an elegant approach to consider magnetic monopoles in spacetime. However, these original works by Reinich and Wheeler only develop the duality formalism to the point of accounting for *source-free* classical electrodynamics in *Abelian* gauge theory. These works do not explain the existence of charge *sources*, either for electromagnetism, or for the weak and strong interactions, the latter two of which are non-Abelian. In fact, when taken together with electromagnetic fields derived from a four-vector potential, duality symmetry appears to require the disappearance of *all* sources, including electric charges.

Beyond resolving the more-than-a-century-old question of magnetic monopoles in Maxwell's electrodynamics, one of the reasons why magnetic monopoles are of interest today is because their appearance in QCD may lead to a better understanding of such phenomena as QCD confinement and superconductivity. Nambu first realized that colored magnetic monopoles, if placed in a QCD vacuum with superconducting properties, would form flux tubes due to Meissner effect, which could help to explain confinement [3] [4]. Later, "'t Hooft became the first to demonstrate the natural appearance of color magnetic monopoles in QCD [4] [5]." Additionally, by introducing a so-called "Abelian gauge," t' Hooft was able to reduce QCD to a QED-like theory, and thus make use of the Dirac Quantization Condition $e \cdot m = n \cdot 2\pi \hbar c$ for unit electric charges e and unit magnetic charges m [6].

Witten, in a lucid article on this whole subject, gets right to the heart of the matter by noting that duality symmetry "seems to be violated when we derive the magnetic field from a vector potential $\mathbf{A}$, with $\mathbf{B} = \nabla \times \mathbf{A}$, while representing the electric field (in a static situation) as the gradient of a scalar." He further goes on to point out that "the vector potential is not just a convenience [but] is needed in 20[th]-century physics for three very good purposes: To write a Schrödinger equation for an electron in a magnetic field. To make it possible to derive Maxwell's equations from a Lagrangian. To write anything at all for non-Abelian gauge theory, which – in our modern understanding of elementary particle physics – is the starting point in describing the strong, weak and electromagnetic interactions." [7] at page 28.

In this paper, and in some subsequent follow-on papers, the author will engage in a detailed exploration of duality in the context of Abelian and non-Abelian gauge theory. As a "recipe" for proceeding, we limit ourselves to consider and combine *only* the following three, simple ingredients: 1) Abelian and Non-Abelian (Yang-Mills) Gauge Theory. 2) Reinich-Wheeler Duality and its underlying Levi-Civita formalism. 3) The Dirac Quantization Condition.

We can hypothesize, because the Reinich-Wheeler Duality symmetry "seems to be violated" [7] by electromagnetism, that this symmetry does not exist in nature. Or, we can hypothesize that this symmetry does exist in nature's underlying Lagrangian, but has somehow become broken (hidden) for the energies at which we observe, and thus try to understand how this occurs. We shall follow the latter path.

By carefully integrating the three well-established and conservatively-grounded ingredients above, and by developing the logical consequences of these three ingredients in combination as far as possible, we can gain not only a better understanding of duality and electric and magnetic charges (a.k.a. monopoles) in their own right, but can arrive at many other interesting results.

---

[*] Maxwell's Equations first appeared in 1873.

## 2. Maxwell's Equations and Gauge Invariance, and Origins of Vanishing Magnetic Monopoles

It is helpful to begin by demonstrating explicitly what Witten means when he says the vector potential is required "to derive Maxwell's equations from a Lagrangian [7]." One starts with a Lagrangian density

$$L_e = i\overline{\psi}_e \gamma^\mu \partial_u \psi_e - \overline{\psi}_e m_e \psi_e \tag{2.1}$$

for an "electric" Dirac spinor wavefunction

$$\psi_e = u_e(p^\mu) e^{-ip^\mu x_\mu} \tag{2.2}$$

of mass $M_e$ and momentum four-vector $p^\mu$, which yields the Dirac field equation:

$$\frac{\partial}{\partial x_\mu}\left(\frac{\partial L}{\partial\left(\frac{\partial \overline{\psi}_e}{\partial x_\mu}\right)}\right) - \left(\frac{\partial L}{\partial(\overline{\psi}_e)}\right) = (i\gamma^\mu \partial_u - m_e)\psi_e = 0. \tag{2.3}$$

(Applied to $\psi_e$, this also yields the adjoint Dirac equation). In the above, we have used $\psi_e$ and $m_e$ rather than the usual $\psi$ and m, because the "e" subscript designates an electric charge which will subsequently be distinguished from a magnetic charge $\psi_m$ when we start to consider magnetic charges. We have also used $L_e$ to denote a Lagrangian for "electrically-charged" Dirac spinors as distinguished from a Lagrangian for magnetic-type charges.

Starting from (2.3), one derives the continuity (charge conservation) equation $J^\mu_{;\mu} = (\overline{\psi}_e \gamma^\mu \psi_e)_{;\mu} = 0$ from which one identifies the electric current four-vector $J^\mu = \overline{\psi}_e \gamma^\mu \psi_e$. Semicolons will be used to designate covariant differentiation with Christoffel symbols wherever the metric tensor $g_{\mu\nu} \neq \eta_{\mu\nu}$.

Demanding the *local* invariance of $L_e$ under an Abelian U(1) gauge transformation of the form $\psi_e \to \psi_e' = e^{ia(x_\mu)}\psi_e$, where $a(x_\mu)$ is a real, non-observable, locally-variable phase angle, we must introduce the gauge-covariant derivative $D_\mu \equiv \partial_\mu + ig_e A_\mu$, where $g_e$ is the running "electric" charge of the U(1) gauge group, which, for electromagnetism, is often designated as $g_e=e$ with charge generator Q=-1 for the electron, while $A_\mu$ is the vector potential which in quantum mechanics is taken to represent the photon. With this, we must amend (2.1) to read:

$$L_e = i\overline{\psi}_e \gamma^\mu (\partial_\mu + ig_e A_\mu)\psi_e - \overline{\psi}_e m_e \psi_e - \tfrac{1}{4} F_{\mu\nu} F^{\mu\nu}, \tag{2.4}$$

where the U(1) field tensor is defined in terms of the vector potential $A_\mu$, as:

$$F_{\mu\nu} = A_{\nu;\mu} - A_{\mu;\nu}. \tag{2.5}$$

It is worth recalling that the term $-\frac{1}{4}F_{\mu\nu}F^{\mu\nu}$ represents the kinetic energy of the photon field $A_\mu$, and is constructed to maintain the local gauge invariance of $L_e$. (See, e.g., [8] at 317.) Equation (2.5) is the tensor formulation of what Witten means when he says that "we derive the magnetic field from a vector potential $\mathbf{A}$, with $\mathbf{B} = \nabla \times \mathbf{A}$, while representing the electric field (in a static situation) as the gradient of a scalar."

It is customary at times to integrate the running charge $g_e$ and electric charge generator $Q_e$ into the current four-vector $J^\mu = g_e \overline{\psi}_e \gamma^\mu Q_e \psi_e$, and at other times to write just $J^\mu = \overline{\psi}_e \gamma^\mu \psi_e$ or $J^\mu = \overline{\psi}_e \gamma^\mu Q_e \psi_e$. If we write $J^\mu = g_e \overline{\psi}_e \gamma^\mu Q_e \psi_e$, the field equation derived from the gauge field terms of Lagrangian (2.4), $L_{e(gauge)} = -\overline{\psi}_e \gamma^\mu Q_e \psi_e g_e A_\mu - \frac{1}{4}F_{\mu\nu}F^{\mu\nu} = -J^\mu A_\mu - \frac{1}{4}F_{\mu\nu}F^{\mu\nu}$, is:

$$\frac{\partial}{\partial x_\mu}\left(\frac{\partial L_{e(gauge)}}{\partial(\frac{\partial A_\nu}{\partial x_\mu})}\right) - \frac{\partial L_{e(gauge)}}{\partial A_\nu} = -\partial_\mu F^{\mu\nu} + J^\mu = 0 \tag{2.6}$$

This field equation, of course, is Maxwell's equation for an electric charge:

$$J^\nu = F^{\mu\nu}{}_{;\mu}. \tag{2.7}$$

This is what is meant by it is "possible to derive Maxwell's equations from a Lagrangian." [7]

Of course, this is only one of two Maxwell's equations in tensor formulation. Now, let us turn to the other Maxwell's equation – the "magnetic equation" – given by:

$$P_{\tau\sigma\nu} \equiv F_{\tau\sigma;\nu} + F_{\sigma\nu;\tau} + F_{\nu\tau;\sigma} = 0 \tag{2.8}$$

We have here *defined* a third-rank antisymmetric tensor $P_{\tau\sigma\nu}$. This tensor $P_{\tau\sigma\nu}$ is equal to zero, not for any inherent reason, but because equation (2.5) for field tensor $F_{\mu\nu} = A_{\nu;\mu} - A_{\mu;\nu}$, substituted into (2.8), identically reduces the antisymmetric combination of terms $F_{\tau\sigma;\nu} + F_{\sigma\nu;\tau} + F_{\nu\tau;\sigma}$ to zero. Absent a field tensor of the form $F_{\mu\nu} = A_{\nu;\mu} - A_{\mu;\nu}$, there is no *a priori* reason why equation (2.8) must equal zero. The fact that equation (2.8) is equal to zero when $F_{\mu\nu} = A_{\nu;\mu} - A_{\mu;\nu}$ is connected, of course, to the fact that we do not seem to observe magnetic monopoles in nature. To further explore this apparent vanishing of magnetic monopoles, we turn to Reinich-Wheeler Duality.

## 3. The Magnetic Monopole Lagrangian, and Construction of an Unbroken, Duality-Invariant Lagrangian

Reinich-Wheeler Duality is based on the Levi-Civita formalism (see [9] at pages 87-89) in which the "dual" $*A^{\sigma\tau}$ of any second-rank antisymmetric tensor $A^{\sigma\tau}$ in four-space $\mathcal{R}^4$ is constructed according to the *discrete* transformation $*A^{\sigma\tau} \equiv \frac{1}{2!}\varepsilon^{\delta\gamma\sigma\tau}A_{\delta\gamma}$.[#] It is a simple exercise to verify that the duality operator * acts similarly to the scalar $i^2 = -1$ insofar as two successive applications of this operator, **=-1, carry $A^{\sigma\tau}$ into $-A^{\sigma\tau}$. In section 5 we shall begin to consider *continuous, local* duality transformations through a duality space defined by a complexion angle $\alpha(x^\mu)$, as reviewed, for example, in [9] at page 108. But to start, we examine only discrete duality transformations.

The Levi-Civita formalism can be used to derive the mathematical identity

$$\tfrac{1}{2} A^{\sigma\tau}\left(B_{\tau\sigma;\nu} + B_{\sigma\nu;\tau} + B_{\nu\tau;\sigma}\right) - *A_{\nu\sigma} * B^{\tau\sigma}{}_{;\tau} = 0, \tag{3.1}$$

which applies to *any* two antisymmetric tensors $A^{\sigma\tau}$ and $B^{\tau\sigma}$ in $\mathfrak{R}^4$, and which will be of great importance in the development to follow here. If the reader is not familiar with this formalism, it is a good exercise to derive (3.1) using $*A^{\sigma\tau} \equiv \tfrac{1}{2!}\varepsilon^{\delta\gamma\sigma\tau}A_{\delta\gamma}$ and $\varepsilon^{\sigma\tau\mu\nu}\varepsilon_{\delta\gamma\mu\nu} = -\delta^{\sigma\tau\mu\nu}{}_{\delta\gamma\mu\nu} = -2!\delta^{\sigma\tau}{}_{\delta\gamma}$, the latter of which contains the minus sign on account of the signature of the metric tensor $g_{\mu\nu} \to \eta_{\mu\nu} = (1,-1,-1,-1)$ in local geodesic coordinates.

If we write identity (3.1) specifically in terms of the field tensor $F^{\sigma\tau}$, that is, if we set $A^{\sigma\tau} = B^{\sigma\tau} = F^{\sigma\tau}$, then (3.1) becomes ([2] at page 251, note 22):

$$\tfrac{1}{2} F^{\sigma\tau}\left(F_{\tau\sigma;\nu} + F_{\sigma\nu;\tau} + F_{\nu\tau;\sigma}\right) - *F_{\nu\sigma} * F^{\tau\sigma}{}_{;\tau} = 0, \tag{3.2}$$

where $*F^{\sigma\tau} \equiv \tfrac{1}{2!}\varepsilon^{\delta\gamma\sigma\tau}F_{\delta\gamma}$. If we apply the duality operator * to each of the $F^{\sigma\tau}$ in (3.2), then, since **=-1, one can also derive the related, duality-rotated identity:

$$\tfrac{1}{2} *F^{\sigma\tau}\left(*F_{\tau\sigma;\nu} + *F_{\sigma\nu;\tau} + *F_{\nu\tau;\sigma}\right) - F_{\nu\sigma} F^{\tau\sigma}{}_{;\tau} = 0. \tag{3.3}$$

Now, in the same way that the electric charge current $J^\mu$ is related to $F^{\mu\nu}$ according to Maxwell's equation (2.7), $J^\nu = F^{\mu\nu}{}_{;\mu}$, a "magnetic monopole" current $P^\mu$ is *defined* in relation to $*F^{\mu\nu}$ as:

$$P^\nu \equiv *F^{\mu\nu}{}_{;\mu}. \tag{3.4}$$

It is important to note that $F^{\mu\nu}{}_{;\mu} = J^\nu$ and $*F^{\mu\nu}{}_{;\mu} = P^\nu$ appear, respectively, in (3.3) and (3.2).

Now, as stated at the outset, we shall work from the hypothesis that duality symmetry does exist in nature's underlying Lagrangian, but has somehow become broken (hidden) for the energies at which we observe. If this is the case, then *until* we have found a way to break the duality symmetry, we need to treat the "magnetic monopole" current $P^\mu$ in exactly the same way as the "electric monopole" current $J^\mu$, and indeed, we will need to fashion a Lagrangian in which duality invariance is completely preserved prior to such symmetry breaking. Therefore, as was the case with $J^\nu = F^{\mu\nu}{}_{;\mu}$, equation (2.7), we will first seek to *derive* $P^\nu \equiv *F^{\mu\nu}{}_{;\mu}$ above from a Lagrangian, rather than simply *define* it as above.

---

[#] In some treatments, the duality operator * is instead expressed as ~, and $*A^{\sigma\tau}$ is written as $\tilde{A}^{\sigma\tau}$. Here, we shall employ the * notation used historically by Reinich and Wheeler in their original treatments of this subject.

To do this, one starts with a "magnetic" Lagrangian (contrast (2.1))

$$L_m = i\bar{\psi}_m \gamma^\mu \partial_u \psi_m - \bar{\psi}_m m_m \psi_m \tag{3.5}$$

for a "magnetic" Dirac spinor wavefunction

$$\psi_m \equiv u_m(p^\mu) e^{-ip^\mu x_\mu} \tag{3.6}$$

of mass $m_m$ and momentum four-vector $p^\mu$, with the subscript "m" designating a "magnetic" charge as distinguished from an "electric" charge. Both $\psi_e$ and $\psi_m$ can properly be regarded as the wavefunctions for "electric monopoles" and "magnetic monopoles," respectively.

As in (2.3), we then derive a Dirac field equation for magnetic monopole $\psi_m$:

$$\frac{\partial}{\partial x_\mu}\left(\frac{\partial L_m}{\partial\left(\frac{\partial \bar{\psi}_m}{\partial x_\mu}\right)}\right) - \left(\frac{\partial L_m}{\partial(\bar{\psi}_m)}\right) = \left(i\gamma^\mu \partial_u - m_m\right)\psi_m = 0, \tag{3.7}$$

and from a continuity equation $P^\mu{}_{;\mu} = \left(\bar{\psi}_m \gamma^\mu \psi_m\right)_{;\mu} = 0$, one identifies the four-vector $P^\mu = \bar{\psi}_m \gamma^\mu \psi_m$ of a conserved magnetic monopole current.

We now demand local invariance of (3.5) under an Abelian U(1) gauge transformation of the form $\psi_m \to \psi_m' = e^{ia(x_\mu)}\psi_m$, and establish another gauge-covariant derivative $D_\mu \equiv \partial_\mu + ig_m M_\mu$, where $g_m$ is a newly-defined running "magnetic" charge, and where $M_\mu$ is a "magnetic potential" vector which represents a magnetic gauge boson. With this, (3.5) becomes:

$$L_m = i\bar{\psi}_m \gamma^\mu \left(\partial_\mu + ig_m M_\mu\right)\psi_m - \bar{\psi}_m m_m \psi_m - \tfrac{1}{4} O_{\mu\nu} O^{\mu\nu}, \tag{3.8}$$

where we have defined a new "other" field tensor

$$*O_{\mu\nu} = M_{\nu;\mu} - M_{\mu;\nu}. \tag{3.9}$$

such that $-\tfrac{1}{4} O_{\mu\nu} O^{\mu\nu}$ is the gauge-invariant kinetic energy of $M_\mu$.

If we now write the magnetic monopole current as $P^\mu = g_m \bar{\psi}_m \gamma^\mu Q_m \psi_m$ where $Q_m$ is a "magnetic" charge generator, then the gauge field part of Lagrangian (3.8) is given by $L_{m(gauge)} = -\bar{\psi}_m \gamma^\mu Q_m \psi_m g_m A_\mu - \tfrac{1}{4} O_{\mu\nu} O^{\mu\nu} = -P^\mu M_\mu - \tfrac{1}{4} O_{\mu\nu} O^{\mu\nu}$, leading to the field equation:

$$\frac{\partial}{\partial x_\mu}\left(\frac{\partial L_{e(gauge)}}{\partial(\frac{\partial M_\nu}{\partial x_\mu})}\right) - \frac{\partial L_{e(gauge)}}{\partial M_\nu} = -O^{\mu\nu}{}_{;\mu} + P^\mu = 0, \qquad (3.10)$$

i.e.,

$$P^\nu = O^{\mu\nu}{}_{;\mu}. \qquad (3.11)$$

Equation (3.11) is derived from a Lagrangian, but does not yet quite represent the magnetic monopole current. If we want (3.11) yield equation (3.4), $P^\nu \equiv *F^{\mu\nu}{}_{;\mu}$, which is the magnetic monopole counterpart to $J^\nu = F^{\mu\nu}{}_{;\mu}$, equation (2.7), then we must set $O^{\mu\nu} \equiv *F^{\mu\nu}$. With this, (3.8) and (3.9) become:

$$L_m = i\overline{\psi}_m \gamma^\mu \left(\partial_\mu + ig_m M_\mu\right)\psi_m - \overline{\psi}_m m_m \psi_m - \tfrac{1}{4} *F_{\mu\nu} *F^{\mu\nu}, \qquad (3.12)$$

and

$$*F_{\mu\nu} = M_{\nu;\mu} - M_{\mu;\nu}, \qquad (3.13)$$

from which now follows Maxwell's magnetic monopole equation:

$$P^\nu \equiv *F^{\mu\nu}{}_{;\mu}. \qquad (3.14)$$

This is how we tie together Maxwell's electric and magnetic equations and derive both from a Lagrangian. The current four vectors for $J^\nu$ and $P^\nu$, therefore, are not independent, but are connected through the duality relation $*F^{\sigma\tau} \equiv \tfrac{1}{2!}\varepsilon^{\delta\gamma\sigma\tau} F_{\delta\gamma}$ between their fields, as well as by identities (3.2) and (3.3) which will be central to further development. Similarly, both $A_\nu$ and $M_\nu$ are related to the same field tensor $F_{\mu\nu}$, via $F_{\mu\nu} = A_{\nu;\mu} - A_{\mu;\nu}$ and $*F_{\mu\nu} = M_{\nu;\mu} - M_{\mu;\nu}$, so that these may be directly related to one another by $M_{\nu;\mu} - M_{\mu;\nu} = \tfrac{1}{2!}\varepsilon_{\delta\gamma\mu\nu}\left(A^{\gamma;\delta} - A^{\delta;\gamma}\right)$.

Having derived both Maxwell's electric monopole equation $J^\nu = F^{\mu\nu}{}_{;\mu}$ and magnetic monopole equation $P^\nu = *F^{\mu\nu}{}_{;\mu}$ from a Lagrangian and linked these together by setting $O^{\mu\nu} \equiv *F^{\mu\nu}$ following (3.11), we now combine the electric and magnetic Lagrangians (2.4) and (3.12), including the charge generators $Q_e$ and $Q_m$, and leaving the running charges $g_e$ and $g_m$ separate from the definitions of $J^\nu$ and $P^\nu$, to yield a single, duality-invariant Lagrangian:

$$\begin{aligned}
L &= \overline{\psi}_e\left(i\gamma^\mu \partial_\mu - m_e\right)\psi_e + \overline{\psi}_m\left(i\gamma^\mu \partial_\mu - m_m\right)\psi_m - g_e\left(\overline{\psi}_e \gamma^\mu Q_e \psi_e\right)A_\mu - g_m\left(\overline{\psi}_m \gamma^\mu Q_m \psi_m\right)M_\mu \\
&\quad - \tfrac{1}{4} F_{\mu\nu} F^{\mu\nu} - \tfrac{1}{4} *F_{\mu\nu} *F^{\mu\nu} \\
&= \overline{\psi}_e\left(i\gamma^\mu \partial_\mu - m_e\right)\psi_e + \overline{\psi}_m\left(i\gamma^\mu \partial_\mu - m_m\right)\psi_m - g_e J^\mu A_\mu - g_m P^\mu M_\mu - \tfrac{1}{4} F_{\mu\nu} F^{\mu\nu} - \tfrac{1}{4} *F_{\mu\nu} *F^{\mu\nu}
\end{aligned} \qquad (3.15)$$

This Lagrangian, by construction, is completely invariant under discrete duality transformations $F^{\mu\nu} \to *F^{\mu\nu}$. Later, we shall break the symmetry of this Lagrangian to see if we can understand why we are able to observe $J^\nu$ but not $P^\nu$, that is, why $J^\nu \neq 0$ but $P^\nu = 0$, thus understanding the (apparent) failure of Maxwell's equation to respect the duality symmetry in (3.15).

Further, we now require that unit charges of the $J^\nu$ and $P^\nu$ are to be related by the Dirac Quantization Condition $e \cdot m = n \cdot 2\pi \hbar c$, which implies that the n=1 *units* of electric and magnetic charge are related by $e \cdot m = 2\pi \hbar c$. Using the notations developed here, this means that the *unit* running electric charges $g_e$ and $g_m$ in (3.15) above are to be related by:

$$g_e \cdot g_m = 2\pi \hbar c. \tag{3.16}$$

Finally, we return to Maxwell's magnetic equation expressed in the more commonly-used third-rank form of (2.8), $P_{\tau\sigma\nu} \equiv F_{\tau\sigma;\nu} + F_{\sigma\nu;\tau} + F_{\nu\tau;\sigma} = 0$. To complete the duality symmetry between electric and magnetic objects, we now *define* a third-rank antisymmetric "electric" tensor along the lines of (2.8), as (the reason for the minus sign will shortly become apparent):

$$J_{\tau\sigma\nu} \equiv -\left(*F_{\tau\sigma;\nu} + *F_{\sigma\nu;\tau} + *F_{\nu\tau;\sigma}\right). \tag{3.17}$$

We now turn to examine these third rank antisymmetric tensors $J_{\tau\sigma\nu}$ and $P_{\tau\sigma\nu}$, and their relation to the first rank vectors $J^\nu$ and $P^\nu$, in detail.

## 4. The Problem of Vanishing Charges in Abelian Gauge Theories with Fields Derived from Potentials

In this section, we highlight what we shall refer to as the "zero charge" or the "vanishing charge" or the "source-free" problem. This problem arises from fact that by using the Abelian field potentials $F_{\mu\nu} = A_{\nu;\mu} - A_{\mu;\nu}$ and $*F_{\mu\nu} = M_{\nu;\mu} - M_{\mu;\nu}$ of (2.5) and (3.13), the tensors $P_{\tau\sigma\nu} = F_{\tau\sigma;\nu} + F_{\sigma\nu;\tau} + F_{\nu\tau;\sigma}$ and $J_{\tau\sigma\nu} \equiv -\left(*F_{\tau\sigma;\nu} + *F_{\sigma\nu;\tau} + *F_{\nu\tau;\sigma}\right)$ of (2.8) and (3.17) identically reduce to zero. $P_{\tau\sigma\nu} = 0$ does not appear to pose a problem, and in fact, $P_{\tau\sigma\nu} = 0$ is just Maxwell's magnetic equation (2.8). But, for $J_{\tau\sigma\nu} = 0$, this is a serious problem, because $J^\sigma = F^{\tau\sigma}{}_{;\tau}$, combined with identity (3.3), $F_{\nu\sigma}F^{\tau\sigma}{}_{;\tau} = \frac{1}{2}*F^{\sigma\tau}\left(*F_{\tau\sigma;\nu} + *F_{\sigma\nu;\tau} + *F_{\nu\tau;\sigma}\right)$ and the use of an Abelian potential $*F_{\mu\nu} = M_{\nu;\mu} - M_{\mu;\nu}$ leads straight to $J^\sigma = 0$, that is, *no electric charge*. We can resolve this either by discarding $*F_{\mu\nu} = M_{\nu;\mu} - M_{\mu;\nu}$ entirely in which case we forego Witten's "three very good purposes" ([7] at page 28) as regards magnetic monopoles. *Or*, we must find a way to have $*F_{\tau\sigma;\nu} + *F_{\sigma\nu;\tau} + *F_{\nu\tau;\sigma} \neq 0$, which means that we must add some additional terms to $*F_{\mu\nu} = M_{\nu;\mu} - M_{\mu;\nu}$ which don't zero out in the antisymmetric combination $*F_{\tau\sigma;\nu} + *F_{\sigma\nu;\tau} + *F_{\nu\tau;\sigma}$. Some further development of the duality is helpful for formally exploring all of this.

Earlier, we made use of the duality relation $*A^{\sigma\tau} \equiv \frac{1}{2!}\varepsilon^{\delta\gamma\sigma\tau}A_{\delta\gamma}$ between two second rank tensors. The Levi-Civita formalism also establishes duality relationships between first rank vectors and third rank antisymmetric tensors given by $*A^\sigma = \frac{1}{3!}\varepsilon^{\alpha\tau\gamma\sigma}A_{\alpha\tau\sigma}$ as well as the inverse relationship $*A_{\tau\sigma\nu} = \varepsilon_{\gamma\tau\sigma\nu} A^{\gamma}$. It is a straightforward exercise to confirm that for first-third rank duality, $**=1$. This is in contrast to $**=-1$ for second-second rank duality (see [9], at pages 87-89). While we use the same symbol $*$ for both types of duality, one can easily tell which duality is employed in any situation just by looking at the rank of the tensors which the $*$ is operating on.

First-third rank duality together with second-second rank duality can be used to derive the further mathematical identity:

$$\tfrac{1}{2} * A^{\sigma\tau} * B_{\tau\sigma\nu} + A_{\nu\sigma} B^{\sigma} = 0. \tag{4.1}$$

Using $**=1$ for the vectors and third rank antisymmetric tensors, together with $**=-1$ for second rank antisymmetric tensors, (4.1) may also be used to yield the duality-related identities $\tfrac{1}{2} * A^{\sigma\tau} B_{\tau\sigma\nu} + A_{\nu\sigma} * B^{\sigma} = 0$, $\tfrac{1}{2} A^{\sigma\tau} * B_{\tau\sigma\nu} - *A_{\nu\sigma} B^{\sigma} = 0$, and $\tfrac{1}{2} A^{\sigma\tau} B_{\tau\sigma\nu} - *A_{\nu\sigma} * B^{\sigma} = 0$. These hold for *any* second rank antisymmetric tensor $A_{\nu\sigma}$, *any* third rank antisymmetric tensor $B_{\tau\sigma\nu}$, and *any* vector $B^{\sigma}$ in $\mathfrak{R}^4$. Derivation of (4.1) and its related identities is a good exercise. It is also good to contrast (4.1) and its related identities with (3.1) and variations of (3.1) which may also be arrived at by duality operations.

Now, with equations (2.7), (2.8), (3.14) and (3.17), we return to write the (3.1)-based identities (3.2) and (3.3) as:

$$*F_{\nu\sigma} P^{\sigma} = *F_{\nu\sigma} * F^{\tau\sigma}{}_{;\tau} = \tfrac{1}{2} F^{\sigma\tau}\left(F_{\tau\sigma;\nu} + F_{\sigma\nu;\tau} + F_{\nu\tau;\sigma}\right) = \tfrac{1}{2} F^{\sigma\tau} P_{\tau\sigma\nu} \tag{4.2}$$

and

$$F_{\nu\sigma} J^{\sigma} = F_{\nu\sigma} F^{\tau\sigma}{}_{;\tau} = \tfrac{1}{2} * F^{\sigma\tau}\left(*F_{\tau\sigma;\nu} + *F_{\sigma\nu;\tau} + *F_{\nu\tau;\sigma}\right) = -\tfrac{1}{2} * F^{\sigma\tau} J_{\tau\sigma\nu}. \tag{4.3}$$

If we now rewrite two of the above (4.1)-based identities as $\tfrac{1}{2} F^{\sigma\tau} P_{\tau\sigma\nu} - *F_{\nu\sigma} * P^{\sigma} = 0$ and $\tfrac{1}{2} F^{\sigma\tau} * P_{\tau\sigma\nu} - *F_{\nu\sigma} P^{\sigma} = 0$, substitute $*P^{\sigma} = \tfrac{1}{3!} \varepsilon^{\alpha\tau\gamma\sigma} P_{\alpha\tau\sigma}$ into the former and $*P_{\tau\sigma\nu} = \varepsilon_{\gamma\tau\sigma\nu} P^{\gamma}$ into the latter, and contrast with (4.2) written as $\tfrac{1}{2} F^{\sigma\tau} P_{\tau\sigma\nu} - *F_{\nu\sigma} P^{\sigma} = 0$, we find that:

$$P_{\tau\sigma\nu} = \varepsilon_{\gamma\tau\sigma\nu} P^{\gamma}; \quad P^{\sigma} = \tfrac{1}{3!} \varepsilon^{\alpha\tau\gamma\sigma} P_{\alpha\tau\gamma}. \tag{4.4}$$

Similarly, if we rewrite the other two (4.1)-based identities as $\tfrac{1}{2} * F^{\sigma\tau} J_{\tau\sigma\nu} + F_{\nu\sigma} * J^{\sigma} = 0$ and $\tfrac{1}{2} * F^{\sigma\tau} * J_{\tau\sigma\nu} + F_{\nu\sigma} J^{\sigma} = 0$, substitute $*J^{\sigma} = \tfrac{1}{3!} \varepsilon^{\alpha\tau\gamma\sigma} J_{\alpha\tau\sigma}$ into the former and $*J_{\tau\sigma\nu} = \varepsilon_{\gamma\tau\sigma\nu} J^{\gamma}$ into the latter, and contrast with (4.3) written as $\tfrac{1}{2} * F^{\sigma\tau} J_{\tau\sigma\nu} + F_{\nu\sigma} J^{\sigma} = 0$, we find that:

$$J_{\tau\sigma\nu} = \varepsilon_{\gamma\tau\sigma\nu} J^{\gamma}; \quad J^{\sigma} = \tfrac{1}{3!} \varepsilon^{\alpha\tau\gamma\sigma} J_{\alpha\tau\gamma}. \tag{4.5}$$

In short, the $P^{\nu}$ and $P_{\tau\sigma\nu}$ are the first / third rank duals of one another, and the $J^{\nu}$ and $J_{\tau\sigma\nu}$ are the first / third rank duals of one another. In contrasting (4.4) and (4.5) to $*P_{\tau\sigma\nu} = \varepsilon_{\gamma\tau\sigma\nu} P^{\gamma}$, $*P^{\sigma} = \tfrac{1}{3!} \varepsilon^{\alpha\tau\gamma\sigma} P_{\alpha\tau\sigma}$, $*J_{\tau\sigma\nu} = \varepsilon_{\gamma\tau\sigma\nu} J^{\gamma}$, and $*J^{\sigma} = \tfrac{1}{3!} \varepsilon^{\alpha\tau\gamma\sigma} J_{\alpha\tau\sigma}$, we find that all of these vector / tensor objects are self-dual, i.e.,

$$J^{\gamma} = *J^{\gamma}; \ J_{\tau\sigma\nu} = *J_{\tau\sigma\nu}; \ P^{\gamma} = *P^{\gamma}; \ P_{\tau\sigma\nu} = *P_{\tau\sigma\nu}. \tag{4.6}$$

The minus sign in the definition (3.17) is what leads to $J^{\gamma} = *J^{\gamma}; \ J_{\tau\sigma\nu} = *J_{\tau\sigma\nu}$, as opposed to what would otherwise have been $J^{\gamma} = -*J^{\gamma}; \ J_{\tau\sigma\nu} = -*J_{\tau\sigma\nu}$ absent the minus sign. Now, we turn back to examine the "zero / vanishing charge" or "source-free" problem, using this formalism.

By virtue of the Abelian electric field potential $F_{\mu\nu} = A_{\nu;\mu} - A_{\mu;\nu}$ in (2.5), and assuming $F^{\mu\nu} \neq 0 \neq *F^{\mu\nu}$ (as we shall do throughout), equation (4.2) reduces identically via $F_{\tau\sigma;\nu} + F_{\sigma\nu;\tau} + F_{\nu\tau;\sigma} = 0$ to:

$$P^{\sigma} = 0; \ P_{\tau\sigma\nu} = 0. \tag{4.7}$$

$P^{\sigma} = 0$ tells us that magnetic monopoles vanish, consistent with observational evidence to date.

Similarly, by virtue of the Abelian magnetic field potential $*F_{\mu\nu} = M_{\nu;\mu} - M_{\mu;\nu}$ in (3.13), equation (4.3) reduces identically via $*F_{\tau\sigma;\nu} + *F_{\sigma\nu;\tau} + *F_{\nu\tau;\sigma} = 0$ to:

$$J^{\sigma} = 0; \ J_{\tau\sigma\nu} = 0. \tag{4.8}$$

which tells us, problematically, that electric charges also vanish, and which we know is *not* true. It is also worth noting from (4.2) and (4.3), or even more succinctly from (4.4) and (4.5), that $P^{\sigma} = 0$ iff $P_{\tau\sigma\nu} = 0$, and $J^{\sigma} = 0$ iff $J_{\tau\sigma\nu} = 0$. And, most importantly, it must be noted that *it is the Abelian potentials $F_{\mu\nu} = A_{\nu;\mu} - A_{\mu;\nu}$ and $*F_{\mu\nu} = M_{\nu;\mu} - M_{\mu;\nu}$ which are directly responsible for reducing all four of $P^{\nu}$, $P_{\tau\sigma\nu}$, $J^{\nu}$ and $J_{\tau\sigma\nu}$ to zero.*

Equations (4.7) and (4.8) together specify Maxwell's *source-free* electrodynamics, which explains our statement in the opening paragraph that the original works by Reinich and Wheeler only develop the duality formalism to the point of accounting for *source-free* classical electrodynamics. For $P_{\tau\sigma\nu} = 0$ and $P^{\sigma} = 0$, this does not appear to present a problem, since this is typically how one expresses the latter part of Witten's statement that "we observe electric charges but not magnetic charges (which are usually called magnetic monopoles)." (see [7] at 28) But, $J_{\tau\sigma\nu} = 0$, and especially $J^{\sigma} = 0$, clearly raises a problem with the former part of Witten's statement, that "we observe electric charges." If we want to move beyond source-free electrodynamics, maintain duality symmetry in Lagrangian (3.15), derive *both* Maxwell's electric and magnetic equations from a Lagrangian, derive electric and magnetic fields from a potential, and "observe electric charges" which are *not* zero, then $J^{\sigma} = 0$ in (4.8) presents the conundrum.

If we take (4.8) at face value, this would mean that we do not observe *electric charges* i.e., electric monopoles, just as we do not observe magnetic monopoles. *This is contradicted by abundant evidence of electric charges*. The fact that in nature, $J^{\sigma} \neq 0$ but $P^{\sigma} = 0$, specifies what is often referred to as the "magnetic monopole problem." In terms of the Reinich-Wheeler duality formalism, we discuss the fact that $J^{\sigma} \neq 0$ but $P^{\sigma} = 0$ by saying that Maxwell's electrodynamics is *not* invariant under duality transformations of the form $F^{\mu\nu} \rightarrow *F^{\mu\nu}$, which take the electric field $\mathbf{E} \rightarrow \mathbf{B}$ and the magnetic field $\mathbf{B} \rightarrow -\mathbf{E}$, (see [7] at 28) and thus take the electric monopole current $J^\mu \to P^\mu$ and the magnetic monopole current $P^\mu \to -J^\mu$. If Maxwell's electrodynamics were to be duality-invariant, we would observe magnetic monopoles equally with electric monopoles, which of course we do not. Or, we would observe neither, which is also not true.

Looking carefully at (4.2) and (4.3), we see that there are really *two* problems that need to be resolved before one can explain why in nature, $J^\sigma \neq 0$ but $P^\sigma = 0$. First, *we must find a way to even allow the sources $J^\sigma$ and $P^\sigma$ to exist as non-zero entities.* That is, we must solve the vanishing-charge problem. At the moment, the identities (4.2) and (4.3) make this impossible, so long as we define the fields $F^{\mu\nu}$ and $*F^{\mu\nu}$ in terms of Abelian potentials according to $F_{\mu\nu} = A_{\nu;\mu} - A_{\mu;\nu}$ and $*F_{\mu\nu} = M_{\nu;\mu} - M_{\mu;\nu}$. Second, once we overcome identities (4.2) and (4.3) and learn how to make at least $J^\sigma$, and possibly $P^\sigma$, non-zero, we need to find a way to break the duality symmetry, so we can understand why we might observe $J^\sigma$ but not $P^\sigma$ at experimentally-obtainable energies.

First, looking at (4.3), the only way to make the electric current $J^\sigma = F^{\tau\sigma}{}_{;\tau}$ non-zero is to make the term $J_{\tau\sigma\nu} \equiv *F_{\tau\sigma;\nu} + *F_{\sigma\nu;\tau} + *F_{\nu\tau;\sigma}$ non zero. Again, $J^\sigma = 0$ iff $J_{\tau\sigma\nu} = 0$. To do this, we can no longer define the field tensor $*F_{\mu\nu}$ as $*F_{\mu\nu} = M_{\nu;\mu} - M_{\mu;\nu}$. Either we entirely discard the idea of deriving this field tensor from the magnetic potential $M_\nu$ and thus abandon Witten's "three very good purposes" when trying to understand magnetic monopoles, or we supplement the equation for $*F_{\mu\nu}$ into the general form:

$$*F_{\mu\nu} = M_{\nu;\mu} - M_{\mu;\nu} + \text{``additional terms that make }*F_{\tau\sigma;\nu} + *F_{\sigma\nu;\tau} + *F_{\nu\tau;\sigma}\text{ non-zero.''} \qquad * \quad (4.9)$$

On the other hand, the field tensor defined as $F_{\mu\nu} = A_{\nu;\mu} - A_{\mu;\nu}$ does zero the term $P_{\tau\sigma\nu} \equiv F_{\tau\sigma;\nu} + F_{\sigma\nu;\tau} + F_{\nu\tau;\sigma}$, hence, $P^\nu = *F^{\mu\nu}{}_{;\mu} = 0$, which is (perhaps) just what we want. In other words, the question now arises, do we just leave $F_{\mu\nu}$ alone, and continue to write:

$$F_{\mu\nu} = A_{\nu;\mu} - A_{\mu;\nu} + \text{``no additional terms''?} \qquad (4.10)$$

Or, do we add terms here as well? And, if we should add terms to (4.9) and (4.10), where ought these terms originate from?

Referring first to (4.9), where might we look to obtain the "additional terms to make $*F_{\tau\sigma;\nu} + *F_{\sigma\nu;\tau} + *F_{\nu\tau;\sigma}$ non-zero," so that $J^\sigma$ can also be made non-zero? One possible approach, which appears quite attractive, is to look at *non-Abelian*, Yang-Mills gauge theories, such as the weak and strong interactions, because the central hallmark of Yang-Mills theories is that they *do* introduce additional terms to the Abelian "baselines" $F_{\mu\nu} = A_{\nu;\mu} - A_{\mu;\nu}$ of (2.5) and $*F_{\mu\nu} \equiv M_{\nu;\mu} - M_{\mu;\nu}$ of (3.13). In fact, these additional terms are at the very heart of non-Abelian Yang-Mills gauge theory. These additional terms are what establish the *non-linear* boson-boson interactions which arise in Yang-Mills theories and make them fundamentally different than Abelian field theories. These additional terms are what bring about the charge "anti-screening" behavior such that "if one creates a quark-antiquark pair an separates them by a distance r, the energy grows linearly with r because of a mysterious 'non-Abelian flux tube' that forms between them." (see [7] at page 30) These terms are what cause non-Abelian couplings to run larger at lower probe energies, and in QCD, are related to the confinement of quarks and gluons. And, in the current context, these additional non-linear terms which come about from Yang Mill gauge theories *do* enable antisymmetric field terms of the general form $F_{\tau\sigma;\nu} + F_{\sigma\nu;\tau} + F_{\nu\tau;\sigma}$ and $*F_{\tau\sigma;\nu} + *F_{\sigma\nu;\tau} + *F_{\nu\tau;\sigma}$ to become non-zero. Once $F_{\tau\sigma;\nu} + F_{\sigma\nu;\tau} + F_{\nu\tau;\sigma}$ and $*F_{\tau\sigma;\nu} + *F_{\sigma\nu;\tau} + *F_{\nu\tau;\sigma}$ are enabled to become non-zero, then it *does* become possible to obtain non-zero (electric and / or magnetic) monopole currents, *even in the face of the mathematical identities (4.2) and (4.3)*. This is in fact a very desirable approach, especially when it comes to fashioning non-zero weak and strong charges, and to arriving at strong magnetic monopoles which might further advance Nambu's efforts to explain superconductivity [3] and Witten's Figure 2, [7] at page 30 by using chromo-magnetic QCD monopoles.

---

* For example, one might consider a term of the form $*F_{\mu\nu} = M_{\nu;\mu} - M_{\mu;\nu} + \varepsilon_{\mu\nu\alpha\beta}\partial^\alpha A^\beta$.

However attractive this may first appear, there is one important drawback here: Maxwell's electrodynamics is an *Abelian* gauge theory. If we rely on *non-Abelian* gauge theories to help $*F_{\tau\sigma;\nu} + *F_{\sigma\nu;\tau} + *F_{\nu\tau;\sigma}$ and possibly $F_{\tau\sigma;\nu} + F_{\sigma\nu;\tau} + F_{\nu\tau;\sigma}$ become non-zero, and therefore rely on non-Abelian theory to make the *Abelian* charge $J^\sigma_{em}$ non-zero, then we must necessarily go beyond the scope of Maxwell's U(1) gauge theory in its own right. That is, by turning to non-Abelian gauge theory to fix a problem with Maxwell's Abelian gauge theory, we will become dependent on various "unified" gauge theories such as the electroweak unification in which the electric charge generator $Q = Y + I^3_L$ sits across two or more interactions, at least one of which is non-Abelian, such as SU(2)$_W$. Or, we may come to depend on even more ambitious unifications, such as, but not limited to, the so-called B-L theories where the weak hypercharge Y is sometimes specified by $Y = \tfrac{1}{2}(B-L) + I^3_R$ and so the electric charge is $Q = Y + I^3_L = \tfrac{1}{2}(B-L) + I^3_R + I^3_L$ (see [10], equations (12.7) and (12.8)). Before taking such steps, we should see if perhaps we can solve the "zero charge / source free" problem for Maxwell's electrodynamics more simply and naturally, solely on the basis of Maxwell's *Abelian* electrodynamics itself, with out resort to *any* gauge groups other than U(1)$_{em}$. This is the approach we shall adopt here. That is, we shall see if there exists a way to resolve the vanishing charge problem solely within the confines of U(1)$_{em}$ without resort to weak or strong or other non-Abelian interactions.

Equations (4.9) and (4.10) force another decision upon us as well. Do we look to add additional terms to (4.10) as well as (4.9), in which case we will also make $P^\sigma$ non zero? Right now, we have $P^\sigma = 0$; $P_{\tau\sigma\nu} = 0$ which is (apparently) desirable, and $J^\sigma = 0$; $J_{\tau\sigma\nu} = 0$ which is clearly not desirable. If we add new terms to (4.10) as well as (4.9), then we will have exactly the opposite problem: we can then obtain $J^\sigma \neq 0$; $J_{\tau\sigma\nu} \neq 0$ which we want, but we will also end up with $P^\sigma \neq 0$; $P_{\tau\sigma\nu} \neq 0$ which we may not want because it would imply the existence of magnetic monopoles which have never been observed. Once again, we confront from a different view, the decades-old question of why duality symmetry seem to be violated by Maxwell's equations.

If we pursue an approach which forces $P^\sigma = 0$ into the underlying Lagrangian, we are effectively casting a die which says that we do not observe magnetic monopoles because they do not exists anywhere in nature, period. If we pursue an approach which allows $P^\sigma \neq 0$, we are casting the reverse die to say that magnetic monopoles *do* exist in nature, but that their existence has been hidden to us so far, because have not yet been able to attain energies at which the existence of $P^\sigma$ becomes experimentally manifest. As we shall see, however, we do not really have a choice here, because the process of rotating the fields $F^{\mu\nu}$, $*F^{\mu\nu}$ through a duality space defined by a *local* complexion angle $\alpha(x^\mu)$, *automatically* introduces new terms into *both* (4.9) and (4.10). These new terms cause the electric current $J^\sigma = \frac{1}{3!}\varepsilon^{\alpha\tau\gamma}J_{\alpha\tau\gamma} = -\frac{1}{3!}\varepsilon^{\alpha\tau\gamma}\left(*F_{\alpha\tau;\gamma} + *F_{\tau\gamma;\alpha} + *F_{\gamma\alpha;\tau}\right)$, as well as the magnetic current $P^\sigma = \frac{1}{3!}\varepsilon^{\alpha\tau\gamma}P_{\alpha\tau\gamma} = \frac{1}{3!}\varepsilon^{\alpha\tau\gamma}\left(F_{\alpha\tau;\gamma} + F_{\tau\gamma;\alpha} + F_{\gamma\alpha;\tau}\right)$ to become non-zero, which in terms of the vanishing charge problem, is clearly desirable. In particular, by using a *local* $\alpha(x^\mu)$, which by definition is able to vary from one point in spacetime to the next, we *will* naturally end up with both $J^\sigma \neq 0$ and $P^\sigma \neq 0$. This will leave us with *no choice* but to break the $J^\sigma$, $P^\sigma$ symmetry to leave $J^\sigma \neq 0$ but $P^\sigma = 0$ at low observation energies. We begin by examining continuous, *global* duality transformations, after which we shall then turn to continuous *local* duality transformations.

## 5. Global Duality Symmetry, and Breaking the Electric and Magnetic Monopole Symmetry

To develop the notion of continuous duality transformations – both global and local – we will make use of the complexion angle $\alpha$ reviewed in [9] at page 108. In particular, we start with the Lagrangian of (3.15), reproduced below:

$$L = \bar{\psi}_e\left(i\gamma^\mu\partial_\mu - m_e\right)\psi_e + \bar{\psi}_m\left(i\gamma^\mu\partial_\mu - m_m\right)\psi_m - g_e J^\mu A_\mu - g_m P^\mu M_\mu - \tfrac{1}{4}F_{\mu\nu}F^{\mu\nu} - \tfrac{1}{4}*F_{\mu\nu}*F^{\mu\nu}, \quad (5.1)$$

and examine how continuous local transformations of the form:

$$F_i^{\sigma\tau} \to F_i^{\sigma\tau\prime} = e^{*\alpha(x^\mu)}F_i^{\sigma\tau}, \quad (5.2)$$

affect L. We also examine how to break the Lagrangian symmetry following such transformations. In this section, we first examine *global* duality transformations under $\alpha$, that is, we examine the special case where $\alpha_{;\mu} = 0$. In the next section, we examine *local* duality transformations under $\alpha(x^\mu)$, that is, we examine the situation where $\alpha_{;\mu} \neq 0$.

Because the second-second rank duality operator $**=-1$, we can use a series expansion to write out the continuous operator $e^{*\alpha}$ of (5.2) in the form $e^{*\alpha} = \cos\alpha + *\sin\alpha$, see [9] at page 108. By operating on $F^{\sigma\tau}$ in the manner of (5.2), we inherently perform the same operation on its dual, i.e., $*F^{\sigma\tau} \to *F^{\sigma\tau\prime} = e^{*\alpha}*F^{\sigma\tau}$. More transparently, if we form $F_i^{\sigma\tau}$ and $*F_i^{\sigma\tau}$ into a "duality doublet," and use $**=-1$, the transformation (5.2) takes on a more transparent form:

$$\begin{pmatrix} F^{\mu\nu} \\ *F^{\mu\nu} \end{pmatrix} \to \begin{pmatrix} F^{\mu\nu\prime} \\ *F^{\mu\nu\prime} \end{pmatrix} = \begin{pmatrix} \cos\alpha & \sin\alpha \\ -\sin\alpha & \cos\alpha \end{pmatrix}\begin{pmatrix} F^{\mu\nu} \\ *F^{\mu\nu} \end{pmatrix}. \quad (5.3)$$

It is an easy exercise to confirm that the final term $-\tfrac{1}{4}F_{\mu\nu}F^{\mu\nu} - \tfrac{1}{4}*F_{\mu\nu}*F^{\mu\nu}$ in (5.1) remains invariant under transformation (5.3). This holds whether $\alpha(x^\mu)$ is global *or* local, because this term contains no derivatives of $\alpha(x^\mu)$ to introduce terms involving the complexion gradient $\alpha_{;\mu}$.

Next, we turn to the currents, which were earlier written in (2.7) and (3.14) as $J^\nu = F^{\mu\nu}{}_{;\mu}$ and $P^\nu \equiv *F^{\mu\nu}{}_{;\mu}$. Now, referring back to (2.6), if we consider the gauge terms $-g_e J^\mu A_\mu - \tfrac{1}{4} F_{\mu\nu} F^{\mu\nu}$ and $-g_m P^\mu M_\mu - \tfrac{1}{4} *F_{\mu\nu} *F^{\mu\nu}$ from (5.1), then we observe that:

$$\frac{\partial}{\partial x_\mu}\left( \frac{\partial\left(-\tfrac{1}{4}F_{\mu\nu}F^{\mu\nu}\right)}{\partial\left(\frac{\partial A_\nu}{\partial x_\mu}\right)} \right) = -\partial_\mu F^{\mu\nu} \text{ and } -\frac{\partial\left(-g_e J^\mu A_\mu\right)}{\partial A_\nu} = g_e J^\nu \qquad (5.4)$$

are the terms which lead to Maxwell's equation for electric charges, written covariantly as:

$$g_e J^\nu = F^{\mu\nu}{}_{;\mu}. \qquad (5.5)$$

A similar line of analysis for magnetic charges leads to:

$$g_m P^\mu = *F^{\mu\nu}{}_{;\mu}. \qquad (5.6)$$

Note, we have specifically designated $g_e$ and $g_m$ *separately* from $J^\mu$ and $P^\mu$, which is based on defining $J^\mu = \overline{\psi}_e \gamma^\mu Q_e \psi_e$ and $P^\mu = \overline{\psi}_m \gamma^\mu Q_m \psi_m$ so as to *not* contain $g_e$ and $g_m$, consistently with the appearance of $g_e$ and $g_m$ in the Lagrangian term $-g_e J^\mu A_\mu - g_m P^\mu M_\mu$. This is in contrast to the alternative convention where this Lagrangian term is sometimes written just as $-J^\mu A_\mu - P^\mu M_\mu$ and the currents are is defined as $J^\mu = g_e \overline{\psi}_e \gamma^\mu Q_e \psi_e$ and $P^\mu = g_m \overline{\psi}_m \gamma^\mu Q_m \psi_m$, in which case Maxwell's equation take on the more familiar form $J^\mu = F^{\mu\nu}{}_{;\mu}$ and $P^\nu \equiv *F^{\mu\nu}{}_{;\mu}$ that has been used elsewhere herein. That is, in $-g_e J^\mu A_\mu - g_m P^\mu M_\mu - \tfrac{1}{4} F_{\mu\nu} F^{\mu\nu} - \tfrac{1}{4} *F_{\mu\nu} *F^{\mu\nu}$ of (5.1), it is the express appearance of $g_e$ and $g_m$ which compels us to write $g_e J^\nu = F^{\mu\nu}{}_{;\mu}$ and $g_m P^\mu = *F^{\mu\nu}{}_{;\mu}$ rather than $J^\nu = F^{\mu\nu}{}_{;\mu}$ and $P^\mu = *F^{\mu\nu}{}_{;\mu}$.

Now, let us consider the behavior of $g_e J^\nu = F^{\mu\nu}{}_{;\mu} \to g_e' J^{\nu'} = F^{\mu\nu'}{}_{;\mu}$ as well as $g_m P^\mu = *F^{\mu\nu}{}_{;\mu} \to g_m' P^{\mu'} = *F^{\mu\nu'}{}_{;\mu}$, under (5.2), (5.3), where $\alpha$ is a global parameter for now. Substituting $F^{\mu\nu'} = \cos\alpha F^{\mu\nu} + \sin\alpha *F^{\mu\nu}$; $*F^{\mu\nu'} = -\sin\alpha F^{\mu\nu} + \cos\alpha *F^{\mu\nu}$ from (5.3) yields $g_e' J^{\nu'} = F^{\mu\nu'}{}_{;\mu} = \left(\cos\alpha F^{\mu\nu} + \sin\alpha *F^{\mu\nu}\right)_{;\mu} = \cos\alpha F^{\mu\nu}{}_{;\mu} + \sin\alpha *F^{\mu\nu}{}_{;\mu} = \cos\alpha g_e J^\nu + \sin\alpha g_m P^\nu$,
$g_m' P^{\nu'} = *F^{\mu\nu'}{}_{;\mu} = \left(-\sin\alpha F^{\mu\nu} + \cos\alpha *F^{\mu\nu}\right)_{;\mu} = -\sin\alpha F^{\mu\nu}{}_\mu + \cos\alpha *F^{\mu\nu}{}_\mu = -\sin\alpha g_e J^\nu + \cos\alpha g_m P^\nu$, or, in compact form:

$$\begin{pmatrix} g_e J^\nu \\ g_m P^\nu \end{pmatrix} \to \begin{pmatrix} g_e' J^{\nu'} \\ g_m' P^{\nu'} \end{pmatrix} = \begin{pmatrix} \cos\alpha & \sin\alpha \\ -\sin\alpha & \cos\alpha \end{pmatrix} \begin{pmatrix} g_e J^\nu \\ g_m P^\nu \end{pmatrix}. \qquad (5.7)$$

In other words, under *global* duality transformations (5.2), the "duality doublets" $\begin{pmatrix} g_e J^\nu \\ g_m P^\nu \end{pmatrix}$ of (5.7) and $\begin{pmatrix} F^{\mu\nu} \\ *F^{\mu\nu} \end{pmatrix}$ of (5.3) transform in precisely the same manner.

To deduce how $A_\mu$ and $M_\mu$ transform, we demand global duality invariance of the term:

$$L_{gJB} = -g_e J^\mu A_\mu - g_m P^\mu M_\mu = -\begin{pmatrix} A_\mu & M_\mu \end{pmatrix}\begin{pmatrix} g_e J^\mu \\ g_m P^\mu \end{pmatrix}$$

$$= -g_e' J^{\mu'} A_\mu' - g_m' P^{\mu'} M_\mu' = -\begin{pmatrix} A_\mu' & M_\mu' \end{pmatrix}\begin{pmatrix} g_e' J^{\mu'} \\ g_m' P^{\mu'} \end{pmatrix} = -\begin{pmatrix} A_\mu' & M_\mu' \end{pmatrix}\begin{pmatrix} \cos\alpha & \sin\alpha \\ -\sin\alpha & \cos\alpha \end{pmatrix}\begin{pmatrix} g_e J^\nu \\ g_m P^\nu \end{pmatrix} = L_{gJB}'$$

.(5.8)

in (5.1). Comparing the final terms on each line of the above yields:

$$\begin{pmatrix} A_\mu \\ M_\mu \end{pmatrix} \to \begin{pmatrix} A_\mu' \\ M_\mu' \end{pmatrix} = \begin{pmatrix} \cos\alpha & \sin\alpha \\ -\sin\alpha & \cos\alpha \end{pmatrix}\begin{pmatrix} A_\mu \\ M_\mu \end{pmatrix}. \tag{5.9}$$

Thus, duality doublet $\begin{pmatrix} A_\mu \\ M_\mu \end{pmatrix}$ also transforms globally in the same way as $\begin{pmatrix} g_e J^\nu \\ g_m P^\nu \end{pmatrix}$ and $\begin{pmatrix} F^{\mu\nu} \\ *F^{\mu\nu} \end{pmatrix}$.

Now, at this point – even without yet having resolved the "zero charge problem" – we actually have enough to consider breaking the symmetry between electric and magnetic charges. In breaking this symmetry, we shall borrow from electroweak theory, and impose what we shall refer to as an "electroweak-like" symmetry breaking condition. Recall that in electroweak theory, the neutral current Lagrangian term is $-g_Y J_Y^\mu B_\mu - g_w J_3^\mu W_{3\mu}$, which contains the weak hypercharge current $J_Y^\mu$, gauge boson $B_\mu$, and hypercharge $g_Y$, and the third component of the weak isospin current $J_3^\mu$, its gauge boson $W_{3\mu}$, and weak charge $g_w$. One then performs a rotation of this Lagrangian term through weak mixing angle $\theta_W$ to arrive at $-g_Y' J_Y^{\mu'} B_\mu' - g_w' J_3^{\mu'} W_{3\mu}'$. As a general rule, $\theta_W$ is not an observable, since its choice is arbitrary. But, as soon as one defines $J_{em}^\mu \equiv J_Y^{\mu'} = J_Y^\mu + I_3^\mu$, one immediately arrives at the relationship $e = \cos\theta_W \cdot g_Y = \sin\theta_W \cdot g_w$, and the weak mixing angle $\theta_W$ takes on a preferred, observable value $\sin^2\theta_W = \dfrac{e^2}{g_w^2} = \dfrac{a_{em}}{a_w}$ which is no longer arbitrary, and which specifies the ratio of the electromagnetic coupling to the weak coupling. In this situation, the weak mixing angle acts similarly to an order parameter which is arbitrary and not observable while one is free to rotate $-g_Y J_Y^\mu B_\mu - g_w J_3^\mu W_{3\mu}$ at will. But, as soon as a symmetry-breaking condition such as $J_{em}^\mu \equiv J_Y^{\mu'} = J_Y^\mu + I_3^\mu$ is imposed, that order parameter is no longer arbitrary and now becomes observable.

Here, the Lagrangian term $-g_e J^\mu A_\mu - g_m P^\mu M_\mu$ is identical in form to the electroweak neutral current Lagrangian term $-g_Y J_Y^\mu B_\mu - g_w J_3^\mu W_{3\mu}$, and the complexion angle $\alpha$ seems to be performing a rotation identical to that performed by $\theta_W$. Here, the analog to the condition $J_{em}^\mu \equiv J_Y^{\mu\prime} = J_Y^\mu + I_3^\mu$ which fixes $\theta_W$ would be $J_{em}^\mu \equiv J^{\mu\prime} = J^\mu + P^\mu$. So we shall see here, what happens if we follow this electroweak analogy all the way through.

From the upper equation of (5.7), we have:

$$g_e' J^{\mu\prime} = \cos\alpha \cdot g_e J^\mu + \sin\alpha \cdot g_m P^\mu. \qquad (5.10)$$

We then *impose* (not derive) the electroweak-like *symmetry-breaking condition*:

$$J_{em}^\mu \equiv J^{\mu\prime} = J^\mu + P^\mu, \qquad (5.11)$$

from which we can immediately identify the relationships:

$$e = g_e' = \cos\alpha \cdot g_e = \sin\alpha \cdot g_m. \qquad (5.12)$$

It will be noted that this bears a clear resemblance to the similar relationships in electroweak theory, $e = \cos\theta_W \cdot g_Y = \sin\theta_W \cdot g_w$. From the lower component of (5.7), we obtain:

$$g_m' P^{\mu\prime} = -\sin\alpha \cdot g_e J^\mu + \cos\alpha \cdot g_m P^\mu. \qquad (5.13)$$

Using (5.11) and (5.12), this may be used to derive:

$$P^{\mu\prime} = P^\mu - \sin^2\alpha \cdot J^{\mu\prime} \qquad (5.14)$$

and

$$g_m' = \frac{g_m}{\cos\alpha} = \frac{g_e}{g_e'} g_m. \qquad (5.15)$$

These, it will be noted, bears a clear resemblance to the electroweak neutral current $J_z^{\mu\prime} = I_3^\mu - \sin^2\theta_W \cdot J_{em}^\mu$, and the z-charge $g_Z = g_w / \cos\theta_W$. From (5.15), and applying (3.16), this means that

$$g_e' \cdot g_m' = g_e \cdot g_m = 2\pi\hbar c, \qquad (5.16)$$

i.e., that the Dirac Quantization Condition (3.16) *is preserved* under the duality transformations (5.2) followed by duality symmetry breaking based on (5.11). In fact, we can now work backwards: if we *require* that the symmetry be broken so as to *preserve* the Dirac Quantization Condition (3.16), i.e., that the duality symmetry be broken such that $g_e' \cdot g_m' = g_e \cdot g_m = 2\pi\hbar c$, then from (5.16), one can work back to the symmetry-breaking condition (5.11). That is, if (5.16) is to be *required*, then it appears as though we *must* impose (5.11).

Of additional interest, equation (5.12) easily yields the relationship:

$$\frac{1}{g_e'^2} = \frac{1}{g_e^2} + \frac{1}{g_m^2} \tag{5.17}$$

which also has a readily-recognizable counterpart in electroweak theory. We rewrite the above, using (3.16) and (5.16), as:

$$g_m'^2 = g_e^2 + g_m^2 = \frac{g_e^2 g_m^2}{g_e'^2} = \frac{(2\pi\hbar c)^2}{g_e'^2}, \tag{5.18}$$

which brings into play for the first time, the term $g_e^2 + g_m^2$ which is the key term of Montonen-Olive duality (see [7] at page 29, [11]), and which by $g_m'^2 = g_e^2 + g_m^2$ is the *duality-rotated unit of magnetic charge $g_m$*.

Now, we turn to the bosons in (5.9), which, it will be noted, mix via $\alpha$ very similarly in form to how the electroweak vector bosons $A^\mu$, $Z^\mu$ mix $B^\mu$ and $W_{3\mu}$ via $\theta_W$. From (5.13), and using (5.12), (5.17) and (5.18), we write:

$$g_m' A_\mu' = \sqrt{g_e^2 + g_m^2} A_\mu' = g_m \cdot A_\mu + g_e \cdot M_\mu \tag{5.19}$$

$$g_m' M_\mu' = \sqrt{g_e^2 + g_m^2} M_\mu' = -g_e \cdot A_\mu + g_m \cdot M_\mu, \tag{5.20}$$

which clearly resemble equations for the massless photon $A^\mu$ and the massive neutral $Z^\mu$ in electroweak theory.

In fact, now we can begin to see what has happened here. By imposing the electroweak-like symmetry-breaking condition (5.11), we have laid the foundation for the duality-rotated photon $A_\mu'$ of (5.9) – which we now identify with the *observed* photon – to be massless. And, we have laid the foundation for the duality rotated $M_\mu'$, which we also take to be *observable*, and which mediates interactions for duality-rotated magnetic monopole currents $P^{\mu'}$, to be *massive*. Following suit, we take the $J^{\mu'}$, which interacts with the duality-rotated $A_\mu'$, to be the *observed* electric monopole current. And, we take $P^{\nu'}$, which interacts with the duality-rotated $M_\mu'$, to be an observable magnetic monopole current that *could* be observed if one were to supply an energy commensurate with the non-zero, presumably large mass of its mediating $M_\mu'$. Finally, we take the *observable* unit charges to be the $g_e'$ and $g_m'$, following duality rotation and symmetry breaking, not the original $g_e$ and $g_m$, and thus in particular identify $g_e'$=e with the unit electric charge.

If all of this is the case, then the fact that we observe electric charges but not magnetic charges can be explained because of the $M_\mu'$ being a very massive vector boson – so massive, in fact, that we simply have not yet been able to observe interactions involving $M_\mu'$, and thus, its magnetic monopole currents $P^{\mu'}$. But, to accord with experiment, the mass of $M_\mu'$ must be in a region that we have not yet explored experimentally. How massive might this be?

Now, the one final step we have *not* yet taken, is to employ the Higgs-Goldstone mechanism to generate a mass for $M_\mu'$. In electroweak theory, this is where the symmetry of the electroweak vacuum is broken to maintain invariance under the U(1)$_{em}$ subgroup such that Goldstone scalars are swallowed up to give a spin-zero longitudinal polarization to the $Z^\mu$ vector boson (as well as the $W^{\pm\mu}$), thus making it massive, while the photon is made massless. For the moment, we defer a detailed exposition of how to apply the Higgs-Goldstone mechanism in the duality vacuum until section 8, pending some further developments that will take place in the next two sections.

For now, let's instead let's cut right to the chase to make an educated guess what the mass of $M_\mu'$ ought to be, especially because of the exact parallels to electroweak theory that we have seen here. For example, in electroweak interactions, the general form for a vector boson mass turns out to be $M = \frac{1}{2}vg$, where for the $W^{\pm\mu}$, $M_W = \frac{1}{2}vg_w$ and for the $Z^\mu$, $M_Z = \frac{1}{2}vg_w' = \frac{1}{2}v\sqrt{g_w^2 + g_Y^2}$, where $g_w \rightarrow g_w'$ following a rotation analogous to (5.8) of the electroweak Lagrangian neutral current terms through $\theta_W$. If we use the Fermi vacuum expectation value $v = 246.220$ GeV[*] [12] just as in electroweak theory,[**] and if vector boson mass magnitudes end up being given by the general form $M = \frac{1}{2}vg$, then in, addition to a massless photon $A_\mu'$ (which allows us to observe electric charge currents $J^{\mu'}$ even at low energy), we arrive at a non-zero mass for the $M_\mu'$ which mediates the interactions of magnetic monopole currents $P^{\mu'}$, given by:

$$\text{Mass}(M^{\mu'}) = \tfrac{1}{2}vg_m' = \tfrac{1}{2}v\sqrt{g_e^2 + g_m^2}. \tag{5.21}$$

To calculate this, we observe that dimensionless running couplings $\alpha$ are typically specified in relation to their associated charges g according to $a = \dfrac{g^2}{4\pi\hbar c}$. So, in light of (5.16), the Dirac Quantization Condition can be written for both primed and unprimed couplings as:

$$2a_m = \frac{1}{2a_e} = \frac{g_m^2}{2\pi\hbar c} = \frac{2\pi\hbar c}{g_e^2}; \quad 2a_m' = \frac{1}{2a_e'} = \frac{g_m'^2}{2\pi\hbar c} = \frac{2\pi\hbar c}{g_e'^2} \tag{5.22}$$

where $a_e' = 1/137.036$ specifies the running strength of the electric charge at low energies (remember that we take $a_e'$, not $a_e$, to be the observable coupling). If, for purposes of "ballpark" estimation, we use $a_e' = 1/137.036$ together with $g_m'^2 = \dfrac{\pi\hbar c}{a_e'}$ and $g_e^2 = 4\pi\hbar c a_e$, (5.21) becomes:

$$\text{Mass}(M^{\mu'}) = \tfrac{1}{2}vg_m' = v\sqrt{\frac{\pi\hbar c}{4a_e'}} = 2.554 \text{ TeV}. \tag{5.23}$$

This is in a range for observing mass events which has not yet been reached experimentally. Of course, in the TeV range, $a_e'$ is already somewhat larger than 1/137.036. For example, in the TOPAZ Collaboration [13], at an average center-of-mass energy of 57.77 GeV, it was observed that $a_e' \sim 1/128.5$. Also, for example, the three-generation graphs in [14] at page 6 suggest $a_e' \sim 1/126$

---

[*] This is related to the Fermi coupling constant according to $\sqrt{2}\, G_F = 1/v^2 = 1/(246.220 \text{ GeV})^2$.
[**] This is an *assumption* which will be examined more closely in section 8.

on the order of 2 TeV. So, we can make a rough estimate that by the time we reach the 2 TeV range, this running of the coupling a$_e$' would bring Mass ($M^{\mu'}$) downward to about 2.35 TeV using a$_e$' ~ 1/126.[*]

With $<\phi>=v$ = Fermi vacuum expectation value,[**] the mass M predicted by Montonen and Olive, [7] at page 30, now becomes associated directly with a mass of about 2.35 TeV for the vector bosons $M^{\mu'}$ which mediate interactions between magnetic monopoles. The photons $A_\mu'$ remain massless. The underlying Lagrangian is invariant under global duality transformations, but the rotation (5.8) and the symmetry-breaking condition (5.11) hide this duality symmetry.

We see therefore, that the symmetry between electric and magnetic currents can be broken strictly by demanding the invariance of Lagrangian (5.1) under *global* duality transformations. Comparing $J_{em}{}^\mu \equiv J^{\mu'} = J^\mu + P^\mu$ of (5.11) with $P^{\mu'} = P^\mu - \sin^2\alpha \cdot J^{\mu'}$ of (5.14), we see that these currents indeed "look different", and their underlying symmetry is now hidden just as that between the $J_{em}{}^\mu$ and $J_Z{}^\mu$ is hidden in electroweak theory. Thus, we have begun to understand a mechanism by which we might observe electric but not magnetic charges at low energies. But, ironically, the "zero charge" problem of section 4 still remains. That is, using global duality symmetry, it appears as though we can break the symmetry between electric and magnetic charges to "hide" the magnetic charges at low energy, but both our electric and magnetic charges are still zero. To solve this vanishing-charge problem, and thus give ourselves non-zero charges *in the first place*, we will need to turn to a *local* duality symmetry. Before we do this however, there are a few more results pertaining to *running couplings* that can be derived strictly from the global duality symmetry as elaborated so far.

## 6. Moving Beyond Perturbation Theory: Parameterizing Running Couplings Through the Complexion Angle

We turn now to take full advantage of the Dirac Quantization Condition (3.16), see also (5.16) and (5.22), which, by giving us a large coupling for every small coupling, has long held out the promise of going beyond perturbation theory to properly deal with large couplings. Equations (3.16), (5.16) and (5.22) can all be used to rewrite (5.12) as:

$$\cos^2\alpha = \frac{g_e'}{g_e} = \frac{a_e'}{a_e}; \quad \sin^2\alpha = \frac{g_e'}{g_m} = \frac{a_e'}{a_m}; \quad \tan^2\alpha = \frac{g_e^2}{g_m^2} = \frac{g_e'^2}{g_m'^2} = \frac{a_e}{a_m} = \frac{a_e'}{a_m'} = (2a_e)^2 \qquad (6.1)$$

The final equation, $\tan^2\alpha = (2a_e)^2$, is arrived at making specific use of the Dirac quantization condition as embodied in (5.23), and illustrates in very clear terms how the running coupling is a function of complexion. This means that a) we should actually be able to numerically answer the question "what is the magnitude of the complexion at low energy where $a_e' = 1/137.036$?"; b) we should be able generally to track the running of all four of the couplings $a_e$, $a_e'$, $a_m$, $a_m'$ in (5.23) as a function of $\alpha$; and, most significantly, c) because of the Dirac Quantization Condition, we should be able to gain an understanding of what happens even when these couplings become *large*, because for every coupling which is small, i.e., which can be reasonably understood using perturbation theory, one *automatically* also has built in, a large "inverse" coupling which cannot be treated with perturbation theory. As Witten [7] notes at page 29, "it is impossible for $a$ and $1/a$ to both be small." So, let is see what else can be derived from this.

---

[*] Observation of this mass could also be used to fine tune calculations of the running of a$_e$' in the 2 TeV range.
[**] Or some other vacuum expectation value if $v$=246.220 GeV turns out to be not suitable.

First, from (5.22) and (6.1), and basic trigonometry, one may derive:

$$2a_e' = \frac{2a_e}{1+(2a_e)^2} \; ; \; 2a_m' = \frac{1+(2a_e)^2}{2a_e} . \tag{6.2}$$

Then, from (6.1),

$$\sin^2 \alpha = \frac{a_e'}{a_m} = 4a_e a_e' = \frac{(2a_e)^2}{1+(2a_e)^2} = \frac{1}{1+(2a_m)^2} \; ; \; \cos^2 \sigma = \frac{a_e'}{a_e} = \frac{1}{1+(2a_e)^2} \tag{6.3}$$

Next, also from (5.22), (6.1):

$$a_e = \tfrac{1}{2}\tan\alpha ; \; a_e' = \tfrac{1}{2}\sin\alpha \cdot \cos\alpha ; \; a_m = \tfrac{1}{2}\cot\alpha ; \; a_m' = \tfrac{1}{2}\sec\alpha \cdot \csc\alpha \tag{6.4}$$

If we then write $a_e' = \tfrac{1}{2}\sin\alpha \cdot \cos\alpha$ above as:

$$(2a_e')^2 = \frac{1}{(2a_m')^2} = \sin^2 \alpha \cdot \cos^2 \alpha = \sin^2 \alpha - \sin^4 \alpha \tag{6.5}$$

we see that this is quadratic in $\sin^2 \alpha$. The solution, combined with (6.3), is:

$$\sin^2 \alpha = \frac{1 \mp \sqrt{1-16a_e'^2}}{2} = \frac{1 \mp \sqrt{1-\frac{1}{a_m'^2}}}{2} = \frac{(2a_e)^2}{1+(2a_e)^2} = \frac{1}{1+(2a_m)^2} \tag{6.6}$$

The positive and negative roots simply interchange the definitions of $\sin^2 \alpha$ and $\cos^2 \alpha$ and thus flip the running curves about $\alpha = \pi/4$, see Figure 1 below.

Thus, given a value for complexion $\alpha$, we can immediately derive the value of all four running couplings via (6.4). Conversely, given any of the four running couplings, we can immediately deduce the complexion $\alpha$ via (6.6), and from that, via (6.4), we can deduce all the other running couplings as well.

So, at low energy, where $a_e' = 1/137.036$, $2a_e' = 1/68.518$, from (6.6), using the negative root, we may deduce:

$$\sin^2 \alpha = 2.131 \times 10^{-4} ; \; \sin\alpha = .0146 = 1/68.511; \; \alpha = .0146 = \pi/215.225 = 0.8363^\circ \tag{6.7}$$

That is, in nature, at low energies, the complexion angler is *not zero*, but is very small, at *less than 1 degree*. In fact, we see from (6.4), that a *zero* complexion would imply that the electromagnetic coupling is zero, so it is the fact that the electromagnetic coupling is small but non-zero goes hand-in-hand with a small, but non-zero complexion angle. It is important to note that the complexion $\alpha$ does not add to the number of parameters needed to describe nature, but rather, is an alternative form in which to express an already-existing natural parameter, namely, the electromagnetic running coupling $a$.

Then, via (6.4) (or (5.22) for $a_m'$), we find that, where $a_e' = 1/137.036$:

$$a_e = \tfrac{1}{2}\tan\alpha = .0073 = 1/137.007; \quad a_m = \tfrac{1}{2}\cot\alpha = 34.252; \quad a_m' = \frac{1}{4a_e'} = \tfrac{1}{2}\sec\alpha \cdot \csc\alpha = 34.259 \quad (6.8)$$

All of these relationships are summarized in Figure 1 on the page following. In particular, not only does Figure 1 show the ordered pair $(\alpha, a_e') = (\pi/215.225, 1/137.036)$ which we observe at low energies, it shows how all of the couplings run as a function of $\alpha$ at higher energies, *including what happens when these couplings grow very large.* Perturbation theory is *not* a limitation here, because these running relationships are based upon the Dirac Quantization condition, for which every small coupling that can be treated with perturbation theory automatically has a large coupling partner which cannot. To be sure, we do not yet know how these couplings run *as a function of probe energy*, but only the general direction of their running: low energy for small $\alpha$, high energy for large $\alpha$. That is, we find that the complexion may be used to parameterize all the running couplings, *small and large.* In seeking to make the all-important connection to probe energy, what is especially worthy of note is how $a_e \cong a_e'$ and $a_m \cong a_m'$ for $\alpha \to 0$, how $a_m \cong a_e'$ and $a_e \cong a_m'$ for $\alpha \to \pi/2$, and how the running of $a_e'$ and $a_m'$ effectively "interpolates" or "bridges" between the extreme $\alpha \to 0$, $\alpha \to \pi/2$ regions where perturbation theory can be applied to two of the four couplings (the small ones) and the middle regions $0 < \alpha < \pi/2$ which we normally do not know how to deal with at all. Thus, by connecting the two extremal regions $\alpha \to 0$, $\alpha \to \pi/2$ which contain couplings that we *do* know how to deal with perturbatively, via these bridges across the region where we do not know how to apply perturbation theory, it may be possible to relate the complexion directly to probe energy and derive an *exact* relationship between running couplings and probe energy, even for very large couplings outside the perturbative region.[*] Also of interest is the central region where $(\alpha, \alpha_e) = (\pi/4, .5)$, because of the obvious symmetry of Figure 1 both vertically and horizontally about this point. Note too, that $a_m' \geq 1$, always, and that $a_e' \leq .25$, always.[**]

## 7. Local Duality Symmetry, Solving the Vanishing Charge Problem, and the Surprising Appearance of SU(2)

Now, we return to resolve the zero-charge problem, by requiring not only global, but *local* duality symmetry. We continue to use Lagrangian (5.1), but now we insist that this Lagrangian remain invariant under *local* duality transformations (5.2). In many instances we shall simply write $\alpha$ rather than $\alpha(x^\mu)$, but it is to be understood at all times from this point forward that $\alpha$ is a *local* parameter.

---

[*] This will be the subject of a separate, subsequent paper by the author.
[**] Finally, it is noted here, and will be developed in a subsequent paper, that figures virtually identical to Figure 1 may be developed for each of the weak and strong interactions. The only real difference is that for these non-Abelian interactions, the low energy region is $\alpha \to \pi/2$ and the high energy region is $\alpha \to 0$, that is, the direction in which the complexion runs relative to probe energy is flipped about $\alpha = \pi/4$, due to charge anti-screening. The interactions then become "unified" at energies where they all have the same complexion (actually, where $|\alpha - \pi/4|$ is identical), i.e., where their running couplings all converge. The apparent separateness of interactions at low energy is manifest by their having different $|\alpha - \pi/4|$ for their complexion, i.e., different values for their "order parameters."

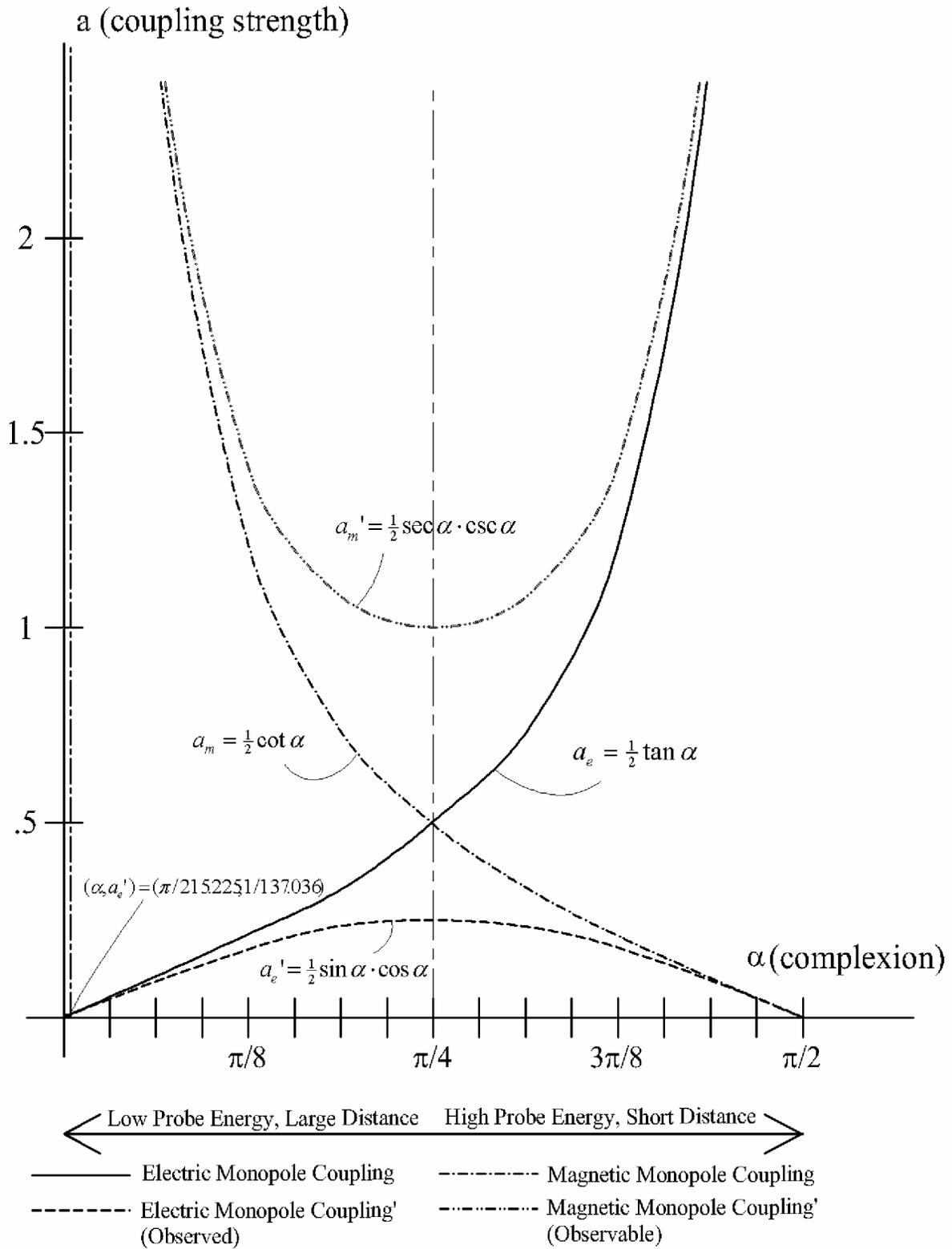

Figure 1: Electromagnetic Running Couplings as a Function of Complexion $\alpha$.

We return first to the global duality transformation (5.9) for *potentials*:

$$\begin{pmatrix} A_\mu \\ M_\mu \end{pmatrix} \to \begin{pmatrix} A_\mu' \\ M_\mu' \end{pmatrix} = \begin{pmatrix} \cos\alpha & \sin\alpha \\ -\sin\alpha & \cos\alpha \end{pmatrix} \begin{pmatrix} A_\mu \\ M_\mu \end{pmatrix}, \tag{7.1}$$

This is also the *local* duality transformation, because there is no derivative defining $\begin{pmatrix} A_\mu \\ M_\mu \end{pmatrix}$.

Now, we turn to the *field strength tensors*. Because of the definitions $F_{\mu\nu} = A_{\nu;\mu} - A_{\mu;\nu}$ and $*F_{\mu\nu} = M_{\nu;\mu} - M_{\mu;\nu}$ in (2.5) and (3.13), these will indeed contain a term which includes the complexion gradient $\alpha_{;\mu}$. For a *local* duality transformation, the field tensors transform as $A_{\nu;\mu} - A_{\mu;\nu} \to A_{\nu;\mu} - A_{\mu;\nu} + M_\nu \alpha_{;\mu} - M_\mu \alpha_{;\nu}$ and $M_{\nu;\mu} - M_{\mu;\nu} \to M_{\nu;\mu} - M_{\mu;\nu} - A_\nu \alpha_{;\mu} + A_\mu \alpha_{;\nu}$, which can be seen by substituting (7.1) into $F_{\mu\nu} = A_{\nu;\mu} - A_{\mu;\nu}$ and $*F_{\mu\nu} = M_{\nu;\mu} - M_{\mu;\nu}$. Of course, we cannot have complexion gradient terms $\alpha_{;\mu}$ appearing in the field tensor, just as in gauge theory, we do not allow phase gradient terms $\Lambda_{;\mu}$ to appear. So, what is the solution? In gauge theory, the usual way to make a Lagrangian invariant under a local, gauge transformation is to introduce a new gauge field which transforms in such a way that the additional phase gradient terms cancel. In U(1)$_{em}$, this gauge field is identified with the photon; in SU(2)$_w$ these become the weak gauge bosons, and in SU(3)$_{QCD}$ these become the gluons. Apparently, to deal with the complexion gradient $\alpha_{;\mu}$, we need to do the same here. So, we shall introduce a new, vector, gauge-like field, which we shall refer to as the "dualon" or "complexon," which we shall designate as $C^\mu$ (The alternative choice, $D^\mu$, could be easily confused with a covariant derivative).

If we pursue this course, and if we *redefine* the fields in terms of the potentials as:

$$\begin{pmatrix} F_{\mu\nu} \\ *F_{\mu\nu} \end{pmatrix} \equiv \begin{pmatrix} A_{\nu;\mu} - A_{\mu;\nu} + M_\nu C_\mu - M_\mu C_\nu \\ M_{\nu;\mu} - M_{\mu;\nu} - A_\nu C_\mu + A_\mu C_\nu \end{pmatrix}, \tag{7.2}$$

then these redefined fields will transform *locally*, according to

$$\begin{pmatrix} F_{\mu\nu}' \\ *F_{\mu\nu}' \end{pmatrix} = \begin{pmatrix} A_{\nu;\mu}' - A_{\mu;\nu}' + M_\nu' C_\mu' - M_\mu' C_\nu' \\ M_{\nu;\mu}' - M_{\mu;\nu}' - A_\nu' C_\mu' + A_\mu' C_\nu' \end{pmatrix}$$
$$= \begin{pmatrix} \cos\alpha & \sin\alpha \\ -\sin\alpha & \cos\alpha \end{pmatrix} \begin{pmatrix} F_{\mu\nu} \\ *F_{\mu\nu} \end{pmatrix} = \begin{pmatrix} \cos\alpha & \sin\alpha \\ -\sin\alpha & \cos\alpha \end{pmatrix} \begin{pmatrix} A_{\nu;\mu} - A_{\mu;\nu} + M_\nu C_\mu - M_\mu C_\nu \\ M_{\nu;\mu} - M_{\mu;\nu} - A_\nu C_\mu + A_\mu C_\nu \end{pmatrix}, \tag{7.3}$$

but <u>if and only if</u> the dualon $C_\mu$ transforms according to:

$$C_\mu \to C_\mu' = C_\mu - \alpha_{;\mu}, \tag{7.4}$$

It is important to note that the local transformation (7.3) is *identical* in form to the global transformation (5.3), and that it is the redefinition (7.2) together with the dualon transformation (7.4) which allows us, locally, to retain these global transformation properties.

Two further things should be noted above. First, it is very interesting that to maintain *local* duality symmetry, we need to add terms of the form $M_\nu C_\mu - M_\mu C_\nu$ and $-A_\nu C_\mu + A_\mu C_\nu$ in (7.2), which look very much like the non-linear interaction terms of a *non-Abelian* gauge interaction. Second, and closely related, it is very interesting that to maintain local duality symmetry, the dualon must transform according to $C_\mu \to C_\mu' = C_\mu - a_{;\mu}$, which looks like an ordinary, U(1) gauge transformation. We will return to both of these points shortly.

Now, we look at the *currents*, and particularly, to the differentiated field tensor terms which transform locally as $F^{\mu\nu}{}_{;\mu} \to F^{\mu\nu'}{}_{;\mu} = F^{\mu\nu}{}_{;\mu} + *F^{\mu\nu}\alpha_{;\mu}$ and $*F^{\mu\nu}{}_{;\mu} \to *F^{\mu\nu'}{}_{;\mu} = *F^{\mu\nu}{}_{;\mu} - F^{\mu\nu}\alpha_{;\mu}$. Here, to cancel the complexion gradients, we need to *redefine* the currents as:

$$\begin{pmatrix} g_e J^\nu \\ g_m P^\nu \end{pmatrix} \equiv \begin{pmatrix} F^{\mu\nu}{}_{;\mu} + *F^{\mu\nu} C_\mu \\ *F^{\mu\nu}{}_{;\mu} - F^{\mu\nu} C_\mu \end{pmatrix}. \tag{7.5}$$

This combination of terms will transform *locally* according to:

$$\begin{pmatrix} g_e' J^{\nu'} \\ g_m' P^{\nu'} \end{pmatrix} = \begin{pmatrix} F^{\mu\nu}{}_{;\mu} + *F^{\mu\nu} C_\mu \\ *F^{\mu\nu}{}_{;\mu} - F^{\mu\nu} C_\mu \end{pmatrix} = \begin{pmatrix} \cos\alpha & \sin\alpha \\ -\sin\alpha & \cos\alpha \end{pmatrix} \begin{pmatrix} g_e J^\nu \\ g_m P^\nu \end{pmatrix}$$
$$= \begin{pmatrix} F^{\mu\nu'}{}_{;\mu} + *F^{\mu\nu'} C_\mu' \\ *F^{\mu\nu'}{}_{;\mu} - F^{\mu\nu'} C_\mu' \end{pmatrix} = \begin{pmatrix} \cos\alpha & \sin\alpha \\ -\sin\alpha & \cos\alpha \end{pmatrix} \begin{pmatrix} g_e J^\nu \\ g_m P^\nu \end{pmatrix} \tag{7.6}$$

<u>if and only if</u> the dualon $C_\mu$ transforms according to $C_\mu \to C_\mu' = C_\mu - \alpha_{;\mu}$, again as in (7.4). This is *identical* in form to the *global* transformation (5.7), and once again, it is $C_\mu \to C_\mu' = C_\mu - a_{;\mu}$, equation (7.4), together with the redefinition (7.5), which enables (7.6), locally, to retain the global transformation properties of (5.7). One can deduce from Lagrangian (5.1), with $J^\mu = \overline{\psi}_e \gamma^\mu Q_e \psi_e$ and $P^\mu = \overline{\psi}_m \gamma^\mu Q_m \psi_m$, that the continuity (conservation) equations $J^\mu{}_{;\mu} = P^\mu{}_{;\mu} = 0$ remain intact.

Now, we return to see whether, by virtue of the "additional terms" we were required to add to the field tensor in (7.2) to maintain a *local* duality symmetry, we may have resolved the zero charge problem, contrast (4.9 and 4.10). For this, we return to the third rank antisymmetric tensors $J_{\tau\sigma\nu} = -(*F_{\tau\sigma;\nu} + *F_{\sigma\nu;\tau} + *F_{\nu\tau;\sigma})$ of (3.17) as well as the $P_{\tau\sigma\nu} = F_{\tau\sigma;\nu} + F_{\sigma\nu;\tau} + F_{\nu\tau;\sigma}$ of (2.8). Into these, we substitute (7.2), and, following the usual cancellation of terms of the form $A_{\tau;\nu;\sigma} - A_{\tau;\sigma;\nu} + A_{\sigma;\tau;\nu} - A_{\sigma;\nu;\tau} + A_{\nu;\sigma;\tau} - A_{\nu;\tau;\sigma} = 0$, $M_{\tau;\nu;\sigma} - M_{\tau;\sigma;\nu} + M_{\sigma;\tau;\nu} - M_{\sigma;\nu;\tau} + M_{\nu;\sigma;\tau} - M_{\nu;\tau;\sigma} = 0$, we now arrive at:

$$\begin{aligned} J_{\tau\sigma\nu} &= -(*F_{\tau\sigma;\nu} + *F_{\sigma\nu;\tau} + *F_{\nu\tau;\sigma}) \\ &= A_\tau(C_{\nu;\sigma} - C_{\sigma;\nu}) + A_\sigma(C_{\tau;\nu} - C_{\nu;\tau}) + A_\nu(C_{\sigma;\tau} - C_{\tau;\sigma}) \\ &+ C_\tau(A_{\sigma;\nu} - A_{\nu;\sigma}) + C_\sigma(A_{\nu;\tau} - A_{\tau;\nu}) + C_\nu(A_{\tau;\sigma} - A_{\sigma;\tau}) \neq 0 \end{aligned} \tag{7.7}$$

and

$$P_{\tau\sigma\nu} = F_{\tau\sigma;\nu} + F_{\sigma\nu;\tau} + F_{\nu\tau;\sigma}$$
$$= M_\tau(C_{\nu;\sigma} - C_{\sigma;\nu}) + M_\sigma(C_{\tau;\nu} - C_{\nu;\tau}) + M_\nu(C_{\sigma;\tau} - C_{\tau;\sigma})$$
$$+ C_\tau(M_{\sigma;\nu} - M_{\nu;\sigma}) + C_\sigma(M_{\nu;\tau} - M_{\tau;\nu}) + C_\nu(M_{\tau;\sigma} - M_{\sigma;\tau}) \neq 0 \qquad (7.8)$$

Because $J^\mu = \frac{1}{3!}\varepsilon^{\tau\sigma\nu\mu}J_{\tau\sigma\nu}$ and $P^\mu = \frac{1}{3!}\varepsilon^{\tau\sigma\nu\mu}P_{\tau\sigma\nu}$, see (4.5) and (4.4), it looks as if we have finally resolved the zero charge problem!

Now, beyond being non-zero, it is important for (7.7) and (7.8) to also remain unchanged under local duality transformations. Looking at the discussion just prior to (7.2), we know that $A_{\nu;\mu} - A_{\mu;\nu} \to A_{\nu;\mu} - A_{\mu;\nu} + M_\nu\alpha_{;\mu} - M_\mu\alpha_{;\nu}$ and $M_{\nu;\mu} - M_{\mu;\nu} \to M_{\nu;\mu} - M_{\mu;\nu} - A_\nu\alpha_{;\mu} + A_\mu\alpha_{;\nu}$ under local duality transformations, and so (7.7) and (7.8) *won't* quite transform the right way as is. At the same time, the term $C_{\nu;\sigma} - C_{\sigma;\nu}$ introduces a new second rank tensor which represents the dualon field, and it is not at all clear how this new field might transform. These are two separate, but closely-related, questions.

For (7.7) and (7.8) to transform properly under local duality, we need throughout to substitute $\begin{pmatrix} A_{\nu;\mu} - A_{\mu;\nu} \\ M_{\nu;\mu} - M_{\mu;\nu} \end{pmatrix} \Rightarrow \begin{pmatrix} A_{\nu;\mu} - A_{\mu;\nu} + M_\nu C_\mu - M_\mu C_\nu \\ M_{\nu;\mu} - M_{\mu;\nu} - A_\nu C_\mu + A_\mu C_\nu \end{pmatrix} = \begin{pmatrix} F_{\mu\nu} \\ *F_{\mu\nu} \end{pmatrix}$, see (7.2), so as to cancel the extra terms $M_\nu\alpha_{;\mu} - M_\mu\alpha_{;\nu}$ and $-A_\nu\alpha_{;\mu} + A_\mu\alpha_{;\nu}$ which arise from local duality transformations.

As regards $C_{\nu;\sigma} - C_{\sigma;\nu}$, how do we define *its* field strength tensor? In particular, looking at (7.2), should $C_{\nu;\sigma} - C_{\sigma;\nu}$ have a term of the form $M_\nu A_\mu - M_\mu A_\nu$ added to it, or should it be left alone? And, if the latter, what should the sign be for *this* term?

First, using (7.4), it is worth noting that the term $C_{\nu;\sigma} - C_{\sigma;\nu}$ by itself, transforms locally as:

$$C'_{\nu;\mu} - C'_{\mu;\nu} = (C_\nu - \alpha_{;\nu})_{;\mu} - (C_\mu - \alpha_{;\mu})_{;\nu} = C_{\nu;\mu} - C_{\mu;\nu} - \alpha_{;\nu;\mu} + \alpha_{;\mu;\nu} = C_{\nu;\mu} - C_{\mu;\nu}, \qquad (7.9)$$

and thus, unlike $\begin{pmatrix} A_{\nu;\mu} - A_{\mu;\nu} \\ M_{\nu;\mu} - M_{\mu;\nu} \end{pmatrix}$, is invariant under local duality transformations. So, in theory, we could leave this term as is. But we also note the transformation properties of the prospective additional term $M_\nu A_\mu - M_\mu A_\nu$, given by (substitute (7.1) into $M_\nu' A_\mu' - M_\mu' A_\nu'$ and reduce):

$$M_\nu' A_\mu' - M_\mu' A_\nu' = M_\nu A_\mu - M_\mu A_\nu, \qquad (7.10)$$

This term is also, by itself, invariant under local duality transformations. Therefore, we do not harm local duality invariance by employing a term $C_{\nu;\sigma} - C_{\sigma;\nu}$ by itself; nor do we harm this invariance by adding or subtracting a term $M_\nu A_\mu - M_\mu A_\nu$. So, what to do?

The answer comes not from demanding local duality invariance, but ironically enough, from demanding local *gauge* invariance. In particular, by adding the terms $M_\nu C_\mu - M_\mu C_\nu$ and $-A_\nu C_\mu + A_\mu C_\nu$ to the field tensors (7.2), we seem to have unwittingly destroyed the U(1) local gauge symmetry we started with, and appear instead to have naturally brought some type of *non-*

*Abelian* gauge symmetry into the picture. That is, local duality symmetry appears to be in conflict with gauge symmetry, or at least, with U(1)$_{em}$ gauge symmetry, but maybe not with non-Abelian gauge symmetry. What has happened here, and what does it portend? Is there some lurking, non-Abelian gauge symmetry that becomes apparent only when we consider local duality symmetry at the same time that we insist on local gauge symmetry?

If one "diagnoses" this from a purely *mathematical* viewpoint, then one can define

$$B_{i\mu} \equiv (A_\mu \quad M_\mu \quad C_\mu), \qquad (7.11)$$

together with an SU(2) field tensor

$$B_{i\mu\nu} \equiv B_{i\nu;\mu} - B_{i\mu;\nu} - \varepsilon_{ijk} B_{j\mu} B_{k\nu}, \qquad (7.12)$$

where $\varepsilon_{ijk}$ are the SU(2) structure constants, and from this, one can recreate precisely the field transformations:

$$\begin{pmatrix} F_{\mu\nu} \\ *F_{\mu\nu} \end{pmatrix} = \begin{pmatrix} A_{\nu;\mu} - A_{\mu;\nu} + M_\nu C_\mu - M_\mu C_\nu \\ M_{\nu;\mu} - M_{\mu;\nu} - A_\nu C_\mu + A_\mu C_\nu \end{pmatrix}, \qquad (7.13)$$

which were required in (7.2) to maintain local duality invariance. From (7.12), we then deduce that

$$C_{\mu\nu} \equiv B_{3\mu\nu} = C_{\nu;\mu} - C_{\mu;\nu} + A_\nu M_\mu - A_\mu M_\nu. \qquad (7.14)$$

Thus, if (7.12) and (7.11) are to apply, we will in fact need to add the term $M_\nu A_\mu - M_\mu A_\nu$ of (7.10) to $C_{\nu;\mu} - C_{;\nu}$ of (7.9), and, (7.12) helps us fix the correct sign for this term. So, again, what does this mean? Are (7.12) and (7.11) valid equations to use here? Is continuous local duality really SU(2) in some unexpected guise? Let's look at the SU(2) transformation properties.

Under an "infinitesimal" SU(2) gauge transformation,

$$B_{i\mu} \to B_{i\mu}' = B_{i\mu}' - \Lambda_{i;\mu} + \varepsilon_{ijk} B^j{}_\mu \Lambda^k, \qquad (7.15)$$

where the $\Lambda_i$, i=1, 2, 3 are the three phase parameters used for SU(2) gauge transformation. For $\Lambda_1 = \Lambda_2 = 0$ and $\Lambda_3 = \alpha$, where $\alpha$ is the complexion, this would seem to indicate that the local duality transformations form a subgroup of SU(2), essentially, the U(1) subgroup of rotations around the third axis in the three-dimensional space of the $B_{i\mu}$. That is, for each of the components of (7.15):

$$A_\mu \to A_\mu' = A_\mu' + M_\mu \Lambda^3 = A_\mu' + M_\mu \alpha, \qquad (7.16)$$

$$M_\mu \to M_\mu' = M_\mu' - A_\mu \Lambda^3 = M_\mu' - A_\mu \alpha, \qquad (7.17)$$

$$C_\mu \to C_\mu' = C_\mu' - \Lambda_{3;\mu} = C_\mu' - \alpha_{;\mu}. \qquad (7.18)$$

If we contrast (7.16) and (7.17), with (7.1) for small $\alpha$, and contrast (7.18) with (7.4), we find, apparently, by insisting upon local duality invariance for $U(1)_{em}$, and by continuing to insist upon local gauge invariance at the same time, we are forced from $U(1)_{em}$ to what looks like an SU(2) gauge symmetry. This SU(2) that is *not* the same as $SU(2)_W$ for weak interactions but originates instead in local duality. And, surprisingly and perhaps quite consequentially, we have found a formal *mathematical* association that can be made between local $U(1)_{em}$ duality symmetry, and local SU(2) gauge symmetry. We shall refer to this as the "SU(2) duality group," or $SU(2)_D$ for short.[*]

With $SU(2)_D$ in hand, returning to (7.7) and (7.8), we now find from (7.14) that we need to substitute $C_{\sigma\nu} = C_{\nu;\sigma} - C_{\sigma;\nu} \Rightarrow C_{\nu;\sigma} - C_{\sigma;\nu} + A_\nu M_\sigma - A_\mu M_\sigma$ throughout as well, to satisfy the local *gauge* symmetry of our newly-discovered $SU(2)_D$ gauge group. Thus, (7.7) and (7.8) reduce to:

$$J_{\tau\sigma\nu} = A_\tau C_{\sigma\nu} + A_\sigma C_{\nu\tau} + A_\nu C_{\tau\sigma} + C_\tau F_{\nu\sigma} + C_\sigma F_{\tau\nu} + C_\nu F_{\sigma\tau} \neq 0 \qquad (7.20)$$

and

$$P_{\tau\sigma\nu} = M_\tau C_{\sigma\nu} + M_\sigma C_{\nu\tau} + M_\nu C_{\tau\sigma} + C_\tau F_{\nu\sigma} + C_\sigma {}^*F_{\tau\nu} + C_\nu {}^*F_{\sigma\tau} \neq 0. \qquad (7.21)$$

It is a straightforward exercise to show that these two tensors now rotate under local duality in the same manner as the currents (7.6), fields (7.3) and potentials (7.1), according to:

$$\begin{pmatrix} J_{\tau\sigma\nu}' \\ P_{\tau\sigma\nu}' \end{pmatrix} = \begin{pmatrix} \cos\alpha & \sin\alpha \\ -\sin\alpha & \cos\alpha \end{pmatrix} \begin{pmatrix} J_{\tau\sigma\nu} \\ P_{\tau\sigma\nu} \end{pmatrix}. \qquad (7.22)$$

Using (4.5) and (4.4) with (7.20) and (7.21), we can now write:

$$J^\mu = \tfrac{1}{3!}\varepsilon^{\tau\sigma\nu\mu} J_{\tau\sigma\nu} = \tfrac{1}{3!}\varepsilon^{\tau\sigma\nu\mu}\left(A_\tau C_{\sigma\nu} + A_\sigma C_{\nu\tau} + A_\nu C_{\tau\sigma} + C_\tau F_{\nu\sigma} + C_\sigma F_{\tau\nu} + C_\nu F_{\sigma\tau}\right) \neq 0 \qquad (7.23)$$

and

$$P^\mu = \tfrac{1}{3!}\varepsilon^{\tau\sigma\nu\mu} P_{\tau\sigma\nu} = \tfrac{1}{3!}\varepsilon^{\tau\sigma\nu\mu}\left(M_\tau C_{\sigma\nu} + M_\sigma C_{\nu\tau} + M_\nu C_{\tau\sigma} + C_\tau F_{\nu\sigma} + C_\sigma {}^*F_{\tau\nu} + C_\nu {}^*F_{\sigma\tau}\right) \neq 0. \qquad (7.24)$$

---

[*] It is then very natural to ask, as a *mathematical* question, is there a similar connection between some expanded type of duality, and larger non-Abelian groups? For example, what would be the nature of a duality that would lead to the larger Yang Mills groups $SU(3)_D$, or $SU(4)_D$, etc.? This is not by any means clear at the moment, but in searching for a phenomenological understanding of the $SU(2)_D$ gauge group that emerges from a local duality symmetry for $U(1)_{em}$, and at the same time in looking to find a mathematical basis (and a physical motivation) for the phenomenology of the fermion generation replication, it is worth exploring if the generation replication and the local non-Abelian Yang Mills gauge groups that naturally emerge from local duality symmetry might in fact support one another once larger groups such as $SU(3)_D$, or $SU(4)_D$ are considered. The author will explore the possibility that local duality symmetry is the source of generation replication, in a subsequent paper.

And so, we have indeed resolved the "zero charge" problem of section 4, using local duality symmetry.*[15]

Returning to the 130-year-old mystery of Maxwell's magnetic monopoles, we find that non-zero electric and magnetic charges and current can indeed be raised into existence by applying local duality symmetry together with local gauge symmetry. The surprise is that in the course of applying local duality symmetry *and* local gauge symmetry *together*, we seem to be compelled to employ a new type of SU(2)$_D$ local gauge symmetry for which a duality rotation through complexion $\alpha$ represents a U(1) rotations around the third axis of SU(2)$_D$, that is, $\Lambda_1 = \Lambda_2 = 0$ and $\Lambda_3 = \alpha$. By breaking this symmetry as outlined in section 5, in a manner very similar to that which is employed in electroweak theory, we also arrive at a possible explanation of why we observe electric charges, but not magnetic charges, at low energies, based on the extreme massiveness of the $M^{\mu'}$ which mediate interactions among magnetic monopoles. And, in combination with the Dirac Quantization Condition, as in section 6, it appears that it may even be possible to consider large running interaction couplings outside the range that can normally be treated with perturbation theory. Now we turn to see if the "educated guess" of the $M^{\mu'}$ mass in equation (5.23) might be borne out by a more formal consideration of the Higgs-Goldstone-type symmetry breaking used to generate mass for the electroweak vector bosons.

## 8. Breaking the Vacuum Symmetry to Generate Boson Mass, and the Vanishing of Magnetic Monopoles in the Low-Energy QED Limit

Using $Q_e$ to designate electric charge and $Q_m$ to designate magnetic charge, let us now postulate two real (not complex) scalars, one of which we designate $\phi_e \equiv |Q_e \neq 0, Q_m = 0 \rangle$ possessing only electric charge and the other $\phi_m \equiv |Q_e = 0, Q_m \neq 0 \rangle$ possessing only magnetic charge. We regard each of these as transforming under a U(1) gauge group, and since these are bosons, they will be specified by the Klein-Gordon Lagrangians:

$$L_e = \tfrac{1}{2}\phi_e^{;\sigma}\phi_{e;\sigma} - \tfrac{1}{2}m^2\phi_e^{\,2} \tag{8.1}$$

$$L_m = \tfrac{1}{2}\phi_m^{;\sigma}\phi_{m;\sigma} - \tfrac{1}{2}m^2\phi_m^{\,2} \tag{8.2}$$

---

* If one uses the photon propagator $A_\sigma = -\dfrac{1}{q^2}J_\sigma$, and writing the current as $J^\mu = \overline{\psi}_e \gamma^\mu Q_e \psi_e$, then the first three terms of (7.20) may be written as:

$$J_{\tau\sigma\nu} = -\frac{1}{q_{(\tau)}^2}\left(\overline{\psi}_{e(\tau)}\gamma_\tau Q_e \psi_{e(\tau)}\right)C_{\sigma\nu} + -\frac{1}{q_{(\sigma)}^2}\left(\overline{\psi}_{e(\sigma)}\gamma_\sigma Q_e \psi_{e(\sigma)}\right)C_{\nu\tau} - \frac{1}{q_{(\nu)}^2}\left(\overline{\psi}_{e(\nu)}\gamma_\nu \psi_{e(\nu)}\right)C_{\tau\sigma} + \ldots,$$

where the indexes in parenthesis associated with the $\psi_e$ and the $q^2$ are labels indicating the spacetime index of their associated currents. This is suggestive that $J_{\tau\sigma\nu}$ may contain three charges within, one from each additive term, and might, in the context of QCD, come to be associated with *baryons*. Because of the dual relationships (4.5) and (4.4), this may even point toward quark – baryon duality, (see, for example, [15]). The author will explore this in a subsequent paper.

If we apply local U(1) gauge transformations $\phi_e \to \phi_e' = e^{ia(x_\mu)}\phi_e$, $\phi_m \to \phi_m' = e^{ia(x_\mu)}\phi_m$, we must then use the gauge-covariant derivative $D_\mu \equiv \partial_\mu + ig_e A_\mu$ for $\phi_e$ and $D_\mu \equiv \partial_\mu + ig_m M_\mu$ for $\phi_m$, so that equations (8.1) and (8.2) become:

$$L_e = \tfrac{1}{2}(\phi_e^{;\sigma} - ig_e A^\sigma \phi_e)(\phi_{e;\sigma} + ig_e A_\sigma \phi_e) - \tfrac{1}{2}m^2 \phi_e^2 = \tfrac{1}{2}\phi_e^{;\sigma}\phi_{e;\sigma} + \tfrac{1}{2}g_e^2 \phi_e A^\sigma A_\sigma \phi_e - \tfrac{1}{2}m^2 \phi_e^2 \qquad (8.3)$$

$$L_m = \tfrac{1}{2}(\phi_m^{;\sigma} - ig_m M^\sigma \phi_m)(\phi_{m;\sigma} + ig_m M_\sigma \phi_m) - \tfrac{1}{2}m^2 \phi_m^2 = \tfrac{1}{2}\phi_m^{;\sigma}\phi_{m;\sigma} + \tfrac{1}{2}g_m^2 \phi_m M^\sigma M_\sigma \phi_m - \tfrac{1}{2}m^2 \phi_m^2. \qquad (8.4)$$

We now form $\phi_e$ and $\phi_m$ into a duality doublet:

$$\Phi \equiv \sqrt{\tfrac{1}{2}}\begin{pmatrix} \phi_e \\ \phi_m \end{pmatrix} \qquad (8.5)$$

transforming according to (contrast, e.g., (7.1)):

$$\begin{pmatrix} \phi_e \\ \phi_m \end{pmatrix} \to \begin{pmatrix} \phi_e' \\ \phi_m' \end{pmatrix} = \begin{pmatrix} \cos\alpha & \sin\alpha \\ -\sin\alpha & \cos\alpha \end{pmatrix}\begin{pmatrix} \phi_e \\ \phi_m \end{pmatrix}. \qquad (8.6)$$

From (8.6) and the U(1) gauge transformations $\phi_e \to \phi_e' = e^{ia(x_\mu)}\phi_e$, $\phi_m \to \phi_m' = e^{ia(x_\mu)}\phi_m$, it is easy to deduce that the quantity $\Phi^\dagger \Phi \to \Phi'^\dagger \Phi' = \tfrac{1}{2}(\phi_e^2 + \phi_m^2)$ is an invariant under local duality and local U(1) gauge transformations.

Now, in electroweak theory, which we continue to use for guidance, one takes mass terms of the form $-\tfrac{1}{2}m^2\phi^2$ and regards them as being the leading terms in a potential of the form $V = \tfrac{1}{2}\mu^2\phi^2 + \tfrac{1}{4}\lambda\phi^4 + \ldots$, finds the minima $\frac{\partial V}{\partial \phi} = (\mu^2 + \lambda\phi^2)\phi = 0$, and assigns these scalars to $\phi^2 = -\mu^2/\lambda = v^2$ where $v$ is a vacuum expectation value. In electroweak theory, one employs the Fermi vacuum expectation value $v=v_F = 246.220$ GeV, see discussion before (5.21), and there are actually four scalar degrees of freedom $\phi^2 = \phi_1^2 + \phi_2^2 + \phi_3^2 + \phi_4^2$.

Here, the term $\tfrac{1}{2}(\phi_e^2 + \phi_m^2)$ involves a pair of scalar fields (two scalar degrees of freedom) which appear to be different than the four scalar fields which appear in the analogous term $\tfrac{1}{2}(\phi_1^2 + \phi_2^2 + \phi_3^2 + \phi_4^2)$ for electroweak theory. So, whereas in electroweak theory one assigns $\langle \phi_1^2 + \phi_2^2 + \phi_3^2 + \phi_4^2 \rangle = v_F^2 = (246.220 \text{ GeV})^2$ in order to match up with experimentally-observed Fermi amplitudes for weak-β decay, we do not know for sure, *a priori*, what vacuum expectation value is suitable for $\tfrac{1}{2}(\phi_e^2 + \phi_m^2)$ of the duality vacuum. This could be the Fermi vacuum, and it could be something else. So, here, we shall follow the general path used in electroweak theory by setting:

$$\Phi^\dagger \Phi \to \Phi'^\dagger \Phi' = \Phi^\dagger \Phi = \tfrac{1}{2}(\phi_e^2 + \phi_m^2) = \tfrac{1}{2}v^2, \qquad (8.7)$$

where $v$ is *some* vacuum expectation value (vev), but we shall also recognize that we do not know for certain whether $v = v_F = 246.220$ GeV is the right vev, or whether some other vev applies here. Of course, since the Fermi vev is related to the Fermi coupling constant $G_F = 1/\sqrt{2}\, v_F^2$, one might suppose that if the $v$ in (8.7) above is something *other* than the Fermi vev, and if the $v$ in (8.7) is also *not* the Planck mass set by the Newton gravitational coupling constant G, then there must be *yet another independent coupling constant* used to fix $v$, which would seem to be an extravagance to be avoided unless absolutely necessary.

In the end, the question whether 246.220 GeV or some other energy is suitable for the vev in (8.7) needs to be established by experimental observation. So, pending experimental evidence one way or the other, we shall for now work on the simple and economical *assumption* that $v = v_F = 246.220$ GeV in (8.7), and will make mass predictions based upon this assumption. For a different vev, the results below can be scaled accordingly.

Now, the question arises, how do we break the symmetry of $\tfrac{1}{2}(\phi_e^2 + \phi_m^2)$ in (8.7), consistently with everything else that has been done herein, and particularly in light of (5.11) which establishes $J_{em}^\mu \equiv J^{\mu\prime} = J^\mu + P^\mu$ as the *observed* electromagnetic current mediated by the observed photon $A_\mu'$ of (5.9) and (5.19) which we wish to make massless. If we look at (8.3) and (8.4), especially the terms $\tfrac{1}{2}g_e^2 \phi_e A^\sigma A_\sigma \phi_e$ and $\tfrac{1}{2}g_m^2 \phi_m M^\sigma M_\sigma \phi_m$, we find that we can combine these terms in the compact form

$$\tfrac{1}{4}\begin{pmatrix}\phi_e' & \phi_m'\end{pmatrix}\begin{pmatrix}g_e' A^{\sigma\prime} \\ g_m' M^{\sigma\prime}\end{pmatrix}\begin{pmatrix}g_e' A_\sigma' & g_m' M_\sigma'\end{pmatrix}\begin{pmatrix}\phi_e' \\ \phi_m'\end{pmatrix}. \tag{8.8}$$

where we have now rotated all of the objects in (8.8) into the primed states via (8.6), since we are associating these with the observables $g_e'$, $A_\sigma'$, $g_m'$, $M_\sigma'$. The above includes an extra factor of ½ because of the $\sqrt{\tfrac{1}{2}}$ in the doublet definition $\Phi \equiv \sqrt{\tfrac{1}{2}}\begin{pmatrix}\phi_e \\ \phi_m\end{pmatrix}$, which also follows customary form. Following this rotation in (8.8), we need to ensure that for the primed states, $\sqrt{\tfrac{1}{2}}\begin{pmatrix}\phi_e' \\ \phi_m'\end{pmatrix} \equiv \begin{pmatrix}|Q_e \neq 0, Q_m = 0\rangle \\ |Q_e = 0, Q_m \neq 0\rangle\end{pmatrix}$, see discussion before (8.1).

Now, to ensure that the photon $A^{\sigma\prime}$ which couples $J_{em}^\mu \equiv J^{\mu\prime} = J^\mu + P^\mu$ remains massless, we will need to ensure that the "duality vacuum" doublet $\begin{pmatrix}\phi_e' \\ \phi_m'\end{pmatrix}$ remains invariant under the U(1) transformations for electric charge $Q_e$. That is, we require (see, e.g., [8], equation (15.16)):

$$\begin{pmatrix}Q_e \neq 0 & 0 \\ 0 & Q_e = 0\end{pmatrix}\begin{pmatrix}\phi_e' \\ \phi_m'\end{pmatrix} = Q_e \phi_e' + Q_e \phi_m' = 0. \tag{8.9}$$

so that:

$$\begin{pmatrix} \phi_e' \\ \phi_m' \end{pmatrix} \rightarrow \begin{pmatrix} \phi_e'' \\ \phi_m'' \end{pmatrix} = e^{ia\begin{pmatrix} Q_e \neq 0 & 0 \\ 0 & Q_e = 0 \end{pmatrix}} \begin{pmatrix} \phi_e' \\ \phi_m' \end{pmatrix} = \begin{pmatrix} \phi_e' \\ \phi_m' \end{pmatrix}. \qquad (8.10)$$

Thus, we must set $\phi_e' = 0$ in the above, so that from (8.7):

$$\begin{pmatrix} \phi_e' \\ \phi_m' \end{pmatrix} = \begin{pmatrix} 0 \\ v \end{pmatrix}. \qquad (8.11)$$

Thus, (8.8) now becomes:

$$\tfrac{1}{4}(0 \quad v) \begin{pmatrix} g_e' A^{\sigma'} \\ g_m' M^{\sigma'} \end{pmatrix} (g_e' A_\sigma' \quad g_m' M_\sigma') \begin{pmatrix} 0 \\ v \end{pmatrix} = 0 \cdot (A^{\sigma'} A_\sigma') + \tfrac{1}{4}(v^2 g_m'^2) M^{\sigma'} M_\sigma'. \qquad (8.12)$$

The scalar degree of freedom from $\phi_e' = 0$ ends up establishing a Higgs field in the usual way, unsatisfying insofar as nothing about that field is readily predicted. But, at the same time, we have revealed a massless photon $A^{\sigma'}$ due to the term $0(A^{\sigma'} A_\sigma')$, as we designed to do. In addition, the scalar degree of freedom from $\phi_m'$ is "swallowed" by the $M_\sigma'$ which now acquires a third, longitudinal polarization and so becomes massive. By comparing (8.12) to the expected Lagrangian term for a vector boson mass m, namely

$$m^2 M^{\sigma'} M_\sigma' = (\tfrac{1}{2} v g_m')^2 M^{\sigma'} M_\sigma', \qquad (8.13)$$

we have also revealed, assuming that $v = v_F = 246.220$ GeV, and using the low energy $a_e' = 1/137.036$ which is expected to run up about 10% at TeV energies, that:

$$\text{Mass } (M^{\mu'}) = \tfrac{1}{2} v g_m' = v \sqrt{\frac{\pi \hbar c}{4 a_e'}} = 2.554 \text{ TeV}, \qquad (8.14)$$

just as in (5.23). Accounting for the running of $a_e' \sim 1/126$ in the 2 TeV range, as noted before, adjusts this mass to about 2.35 TeV. Our earlier "educated guess" seems to be borne out by a detailed consideration of symmetry breaking, and we do indeed find a massive vector boson $M^{\mu'}$ with a mass upwards of 2 TeV which appears to be responsible for our inability to observe magnetic monopoles at low energies.

Finally, let us examine how the results derived here break down to the usual U(1)$_{em}$ symmetry in the low energy limit, where we know that electric monopoles currents $J^\mu$ are observed, but that magnetic monopole currents $P^\mu$ are not. We start with the duality rotated Lagrangian terms from (5.8), namely:

$$L_{gJB}' = -g_e' J^{\mu'} A_\mu' - g_m' P^{\mu'} M_\mu', \qquad (8.15)$$

following the imposition of the symmetry breaking condition $J_{em}{}^{\mu} \equiv J^{\mu\prime} = J^{\mu} + P^{\mu}$ in (5.11).

Because the photon $A_{\mu}'$ is massless, see (8.12), $A_{\mu}'$ will be related to the electric monopole current $J^{\mu\prime} = J^{\mu} + P^{\mu}$ by:

$$A_{\mu}' = \frac{-g_{\mu\lambda}}{p^{\tau}p_{\tau}} J^{\lambda\prime} \equiv \Pi_{\mu\lambda} J^{\lambda\prime}, \tag{8.16}$$

where $\Pi_{\mu\lambda}$ in the above is the propagator for the massless photon.

In contrast, because $M^{\mu\prime}$ is massive, $m = \tfrac{1}{2} v g_m'$, see (8.14), $M^{\mu\prime}$ will be related to the observable magnetic monopole current $P^{\mu\prime} = P^{\mu} - \sin^2 \alpha \cdot J^{\mu\prime}$, see (5.14), by:

$$M_{\mu} = \frac{-g_{\mu\lambda} + p_{\mu}p_{\lambda}/m^2}{p^{\tau}p_{\tau} - m^2} P^{\lambda\prime} \equiv \Pi_{\mu\lambda} P^{\lambda\prime}, \tag{8.17}$$

where $\Pi_{\mu\lambda}$ in the above is the propagator for a massive vector boson, in this case, the massive mediator $M^{\mu\prime}$ of magnetic monopole interactions.

Substituting (8.16) and (8.17) into (8.15), then substituting $m = \tfrac{1}{2} v g_m'$ from (8.14) into the above, and finally using the Dirac Quantization Condition (5.16) written as $g_m' = \dfrac{2\pi \hbar c}{g_e'}$ to eliminate $g_m'$ so that only $g_e^2 = 4\pi \hbar c a_e \overset{\text{low energy}}{=} \dfrac{4\pi \hbar c}{137.036}$ remains (see (5.22)), we now arrive at:

$$\begin{aligned}
L_{gJB}' &= -g_e' J^{\mu\prime} \frac{-g_{\mu\lambda}}{p^{\tau}p_{\tau}} J^{\lambda\prime} - g_m' P^{\mu\prime} \frac{-g_{\mu\lambda} + p_{\mu}p_{\lambda}/m^2}{p^{\tau}p_{\tau} - m^2} P^{\lambda\prime} \\
&= \frac{g_e'}{p^{\tau}p_{\tau}} J^{\mu\prime} J_{\mu}' - g_m' \frac{-g_{\mu\lambda} + p_{\mu}p_{\lambda}/(\tfrac{1}{2} v g_m')^2}{p^{\tau}p_{\tau} - (\tfrac{1}{2} v g_m')^2} P^{\mu\prime} P^{\lambda\prime} \\
&= \frac{g_e'}{p^{\tau}p_{\tau}} J^{\mu\prime} J_{\mu}' - \frac{-g_{\mu\lambda}\left(\dfrac{2\pi\hbar c}{g_e'}\right) + \dfrac{p_{\mu}p_{\lambda}}{\tfrac{1}{4} v^2}\left(\dfrac{g_e'}{2\pi\hbar c}\right)}{p^{\tau}p_{\tau} - \tfrac{1}{4} v^2 \left(\dfrac{2\pi\hbar c}{g_e'}\right)^2} P^{\mu\prime} P^{\lambda\prime}
\end{aligned} \tag{8.18}$$

In the low-energy limit, where $p^{\tau}p_{\tau} \ll \dfrac{\pi}{2} \hbar c v^2$, this reduces to:

$$L_{gJB}' = \frac{g_e'}{p^{\tau}p_{\tau}} J^{\mu\prime} J_{\mu}' - \frac{2 g_e'}{\pi \hbar c v^2} P^{\mu\prime} P_{\mu}' = g_e' \left[ \frac{1}{p^{\tau}p_{\tau}} J^{\mu\prime} J_{\mu}' - \frac{2}{\pi \hbar c v^2} P^{\mu\prime} P_{\mu}' \right]. \tag{8.19}$$

Note that in this low-energy limit, the *ratio* of the magnitude of the $J^{\mu\prime}J_\mu{}'$ to the $P^{\mu\prime}P_\mu{}'$ coefficient is given by $\dfrac{\pi\hbar c}{2}\dfrac{v^2}{p^\tau p_\tau}$, that is, this ratio is determined strictly by the ratio of the duality vev to the probe energy. It is clear that where $p^\tau p_\tau << \dfrac{\pi}{2}\hbar c v^2$, (8.19) reduces to:

$$L_{gJB}{}' \cong \frac{g_e{}'}{p^\tau p_\tau} J^{\mu\prime} J_\mu{}' = -g_e{}' J^{\mu\prime} A_\mu{}'. \tag{8.20}$$

This, of course, is U(1) QED Lagrangian interaction term that we see at low energies, and the magnetic monopole interaction term is gone! This demonstrates in more formal terms why we don't observe magnetic monopoles, but do observe electric charges, at low energies.

To see how greatly the magnetic monopole interactions are suppressed, let us consider a probe energy at the nucleon scale 1 fm ~ 1/.197 GeV, that is, at $p^\tau p_\tau = (.197\text{ GeV})^2$. And, let us continue to make the *assumption* that $\dfrac{\pi}{2}\hbar c v^2 = \dfrac{\pi}{2}\hbar c v_F{}^2 = (308.591\text{ GeV})^2$, i.e., that the duality vev is the Fermi vev, as discussed above. In this situation, the coefficient $\dfrac{1}{p^\tau p_\tau}$ which multiplies $J^{\mu\prime}J_\mu{}'$ in (8.19) is 2.44 x $10^6$ times as large as the coefficient $\dfrac{2}{\pi\hbar c v^2}$ which multiplies $P^{\mu\prime}P_\mu{}'$. That is, at energies near 1fm, which is close to $\Lambda_{QCD}$, the magnetic monopole term of the Lagrangian is suppressed relative to the electric monopole term by a factor of 2.44 million. Although small, this is not zero, and so it may well be possible to observe these effects with suitable accuracy, assuming the duality vev is not actually larger than 246.220 in which case the suppression would be even greater. At the Compton wavelength, ~386 fm, the suppression is 3.64 x $10^{11}$, that is, 364 billion. And, at the Bohr radius ~ 5.29 x $10^4$ fm, the suppression factor is 6.84 x $10^{15}$. Of course, for other assumptions (or perhaps, other experimental findings) about the real magnitude of *v*, these numbers will change, but it is clear that unless *v* is *significantly smaller* than 246.220 GeV, one will indeed continue to find that the term $g_m{}' P^{\mu\prime} M_\mu{}'$ remains very highly suppressed in magnitude relative to $g_e{}' J^{\mu\prime} A_\mu{}'$ for all probe energies which do not penetrate the nucleon, and even for some energies which do penetrate, but not very deeply. Indeed, to trigger a $g_m{}' P^{\mu\prime} M_\mu{}'$ interaction, one must produce a vector boson mass of over 2 TeV if $v = v_F = 246.220$ GeV, and even higher if $v_F \neq v > 246.220$ GeV. Present-day experimentation has not yet been capable of producing mass events over 2TeV, but it is starting to get close.

## 9. Conclusion

In conclusion, we have shown, if one requires continuous duality transformations to hold locally, together with continuing to insist on local gauge symmetry, that one arrives naturally at a SU(2)$_D$ duality gauge group structure which also resolves the "source-free" problem of vanishing currents. We have further shown how to break the duality symmetry between these now-non-vanishing electric and magnetic monopoles so that magnetic monopole interactions are very highly suppressed relative to those of electric monopoles at low energies, and so as to yield a mass of about 2.35 TeV for the vector bosons which mediate magnetic monopole interactions if the duality vacuum vev should turn out to be the same as the 246.220 GeV vev of electroweak theory. For other vev, these results scale accordingly (note that the suppression of magnetic monopole interactions varies with the *square* of the vev, see (8.19)). On the atomic scale where electrons dominate, magnetic monopole interactions are effectively zero in relation to electric charge interactions, as is clearly observed.

The most important next step, which the author plans to take in a follow up paper in the near future, is to develop the SU(2)$_D$ duality group that surprisingly emerged here, but starting with gauge theory rather than duality theory. In equation (7.12), we came across SU(2)$_D$ "in the middle," that is, at the level of the field tensors, and used this to "diagnose" the presence of an underlying SU(2)$_D$ group. Now, it is important as the next step, to fully examine this group "from the top." To do this, one would start with the Dirac equation $(i\gamma^\mu \partial_u - m_e)\Psi = 0$ operating on an SU(2)$_D$ doublet of Dirac spinors, develop the Lagrangian while imposing local gauge symmetry, establish a duality relationship between the fields $B_{1\mu\nu}$ and $B_{2\mu\nu}$ of equation (7.12), (that is, set $B_{2\mu\nu} = *B_{1\mu\nu}$), establish local duality rotations through complexion $\alpha$, impose the Dirac Quantization Condition, and then, in particular, take a close look at how all of this impacts on the electric and magnetic monopoles themselves, regarded as fermions which are Dirac spinors.

At this point in time, experimental physicists are just starting to seriously probe TeV-energy phenomena, for example, at the Tevatron. If a massive vector boson $M^{\mu\prime}$ can be observed near about 2.35 TeV, and if this vector boson has all the properties of the photon $A_\mu'$ with the single exception that it is massive rather than massless and thus has a short range, then this may indirectly confirm the results herein as a possible solution to the 130-year-old mystery of Maxwell's magnetic monopoles, and it would confirm a possible relationship between the duality vacuum and the Fermi vacuum. If we can also observe Fermions $\psi_m$ similar to electrons $\psi_e$, whose interactions are mediated by these $M^{\mu\prime}$, this might then be taken as direct evidence of the existence of the magnetic monopoles $\psi_m$.[*] [16], [17], [18], [19], [20], [21], [22], [23], [24]

## Acknowledgements

The author wishes to thank and acknowledge Dr. Björn Feuerbacher for reviewing over half a dozen drafts of this paper, in detail, with great energy and penetrating insight. In particular, it was Dr. Feuerbacher who urged the author to try to resolve the zero-charge problem using only Maxwell's U(1) Abelian gauge theory, without relying on non-Abelian weak and / or strong interactions. Dr. Feuerbacher also came up with the name "dualon," and more importantly, helped overcome the author's initial reluctance to introduce this new, gauge-like field which, in the end, was integral in solving the zero charge without using non-Abelian gauge theory and helped bring about the surprising appearance of SU(2)$_D$ in section 7.

---

[*] It is only noted here, but will be developed in a separate paper, that in weak interactions, the analog to the 2+TeV magnetic monopole mediators is a 1.231 TeV vector boson to mediate interactions of weak magnetic monopoles, based on low energy electromagnetic and weak couplings which will run somewhat from their low energy value when probed in the TeV range. The strong interaction of course runs dramatically, but at the $\pi/4$ symmetry point in Figure 1, the mediators of interactions for chromomagnetic monopoles appear to have a mass of 436 GeV. These latter weak and strong massive vector boson masses are not far from candidate mass events recently-observed at the Tevatron, see references [16] through [24].

The author wishes also to thank and acknowledge Dr. Andrej E. Inopin for many hours of consultation and encouragement during the early development of this work while the author was still relying on non-Abelian gauge theory to resolve the zero charge problem, and for his knack for always being able to find just the right article for me to consider at just the right time. Thanks and appreciation to Mr. Fred Diether, for numerous helpful suggestions to improve the readability of the paper, and for his general advice and counsel. Finally, I'd like to acknowledge and thank my teacher, Dr. Hans C. Ohanian. Over twenty years ago he handed me a slip of paper with two references written on it, and suggested that I see what I can make of them. I did. These were: Reinich [1], and Wheeler [2].

Notwithstanding this valuable input from others, the author, of course, is fully responsible for any errors of form or substance.